\documentclass[preprint,12pt]{elsarticle}

\usepackage{amssymb}
\usepackage{amsmath}
\usepackage{booktabs}
\usepackage{siunitx} 
\usepackage{multirow}
\usepackage{tabu}
\usepackage{subcaption}
\usepackage{algorithm}
\usepackage{algpseudocode}
\usepackage{stmaryrd}
\usepackage{graphicx}
\usepackage{fancyhdr}
\usepackage{xcolor}
\usepackage{float}
\usepackage{mathrsfs}
\usepackage[normalem]{ulem}
\usepackage{amsthm}

\journal{ArXiv}

\begin{document}

\begin{frontmatter}



\title{Physics-informed Neural Networks for Heterogeneous Poroelastic Media}

\author[inst1]{Sumanta Roy}
\author[inst1]{Chandrasekhar Annavarapu\corref{cor1}}
\ead{annavarapuc@civil.iitm.ac.in}
\affiliation[inst1]{organization={Department of Civil Engineering},
            addressline={Indian Institute of Technology Madras}, 
            city={Chennai},
            postcode={600036}, 
            state={Tamil Nadu},
            country={India}}

\author[inst2]{Pratanu Roy}
\affiliation[inst2]{organization={Atmospheric, Earth and Energy Division},
            addressline={Lawrence Livermore National Laboratory}, 
            city={Livermore},
            postcode={94551}, 
            state={California},
            country={United States}}
            
\author[inst3]{Dakshina Murthy Valiveti}
\affiliation[inst3]{organization={Engineering and Computational Physics, Energy Sciences and Research},
            addressline={ExxonMobil Technology and Engineering},
            city={Spring},
            postcode={77389}, 
            state={Texas},
            country={United States}}
            
\cortext[cor1]{Corresponding author}

\begin{abstract}
\sloppy
This study presents a novel physics-informed neural network (PINN) framework for modeling poroelasticity in heterogeneous media with material interfaces. The approach introduces a composite neural network (CoNN) where separate neural networks predict displacement and pressure variables for each material. While sharing identical activation functions, these networks are independently trained for all other parameters. To address challenges posed by heterogeneous material interfaces, the CoNN is integrated with the Interface-PINNs or I-PINNs framework (Sarma et al. 2024; https://dx.doi.org/10.1016/j.cma.2024.117135), allowing different activation functions across material interfaces. This ensures accurate approximation of discontinuous solution fields and gradients. Performance and accuracy of this combined architecture were evaluated against the conventional PINNs approach, a single neural network (SNN) architecture, and the eXtended PINNs (XPINNs) framework through two one-dimensional benchmark examples with discontinuous material properties. The results show that the proposed CoNN with I-PINNs architecture achieves an RMSE that is two orders of magnitude better than the conventional PINNs approach and is at least 40 times faster than the SNN framework. Compared to XPINNs, the proposed method achieves an RMSE at least one order of magnitude better and is 40\% faster.
\end{abstract}




\end{frontmatter}

\graphicspath{{Figures/}}

\section{Introduction}\label{sec:introduction}

Coupled-field problems are ubiquitous in the field of subsurface mechanics. Various geothermal and subsurface energy applications are governed by the coupled phenomenon of fluid flow dynamics and solid deformation within poroelastic media. This encompasses activities like hydrocarbon retrieval from oil and gas reservoirs~\cite{fielding1987application}, geological storage of carbon dioxide~\cite{morris2009injection}, and harnessing energy from subsurface thermal systems~\cite{bloemendal2014achieve}. In reservoir engineering~\cite{settari1998coupled,menin2008mechanism}, these interactions are key in determining the state of subsurface fluids and gases, the performance of production wells, and the deformation of the porous subsurface media. Furthermore, the inherent heterogeneity characterizing these subsurface systems adds a layer of complexity. Heterogeneities, in terms of material interfaces, may stem from the natural stratification of the earth’s crust, the presence of naturally occurring faults, or induced variations due to processes such as hydraulic fracturing~\cite{lecampion2018numerical}. The presence of these interfaces introduces weak or strong discontinuities in these coupled output fields, and their gradients. Therefore, a thorough investigation and comprehension of subsurface heterogeneity in terms of material interfaces and the intricate hydro-poromechanical processes emerge as imperative aspects of studying subsurface applications.

Computational modeling is often necessary to supplement these theoretical and experimental investigations to gain fundamental insight into the coupled physical processes of flow and deformation. These computational models can be broadly categorized as either physics-based or data-driven. Physics-based models involve the solution of coupled partial differential equations (PDEs). These coupled PDEs are solved numerically, often employing methods like the finite element method~\cite{ammosov2022generalized,adia2021combined}, finite volume method~\cite{sokolova2019multiscale}, or finite difference method~\cite{ahmad2022numerical}, which provide approximate solutions to the PDEs on an underlying grid. However, generating grids that conform to subsurface features (such as material interfaces, faults, and fractures) is a time-consuming pre-processing step. For instance, in several reservoir engineering applications, the cost of grid generation is comparable to (or exceeds) the cost of analysis~\cite{valiveti2023grid}. This makes computational modeling intractable for realistic field-scale applications. By contrast, data-driven models employ machine learning (ML), particularly artificial neural networks (ANNs), to predict the coupled behavior. Unlike physics-based modeling techniques, these approaches do not rely on an underlying grid for approximation. Instead, they utilize large training datasets and ANNs to predict the physical system’s behavior. However, the precision of these models is highly dependent on the quality and quantity of the training data. This training dataset is often generated through physics-based simulators or experimental methods. Unfortunately, the availability of such training data in subsurface systems is often limited, which restricts the applicability of purely data-driven modeling approaches in poroelasticity. Moreover, data-driven models often perform poorly during extrapolation, i.e. outside the range of training data set.

Physics-informed neural networks (PINNs), a recently developed hybrid approach~\cite{raissi2019physics}, combines aspects of both physics-based and data-driven modeling. On the one hand, PINNs are not grid-based, thereby eliminating the need for conforming meshes. On the other hand, PINNs leverage fundamental physical laws, including governing PDEs, initial and boundary conditions, and data, to construct a loss functional and thereby do not require extensive training datasets. PINNs approximate unknown field variables using feedforward neural networks to minimize the loss functionals constructed from governing PDEs, initial and boundary conditions at randomly selected collocation points across the domain. PINNs have been successfully applied to solve problems in various domains, including solid mechanics~\cite{haghighat2021physics,zhang2022analyses}, fluid mechanics~\cite{cai2021physicsfluid,wessels2020neural}, heat transfer~\cite{cai2021physics,jalili2024physics}, inverse problems~\cite{sarkar1ause,chen2020physics} to name a few.

Despite these advances, the application of PINNs to poroelasticity has been limited. Notably, Haghighat et al.~\cite{haghighat2022physics} introduced a sequential training approach based on stress-split algorithms from poromechanics, solving three benchmark homogeneous problems: Mandel’s consolidation problem, Barry–Mercer’s injection-production problem, and a two-phase drainage problem. However, their framework produced poor approximations as the coupling between solid deformation and fluid flow increased.  Bekele~\cite{bekele2020physics} proposed a PINNs framework to solve Barry and Mercer’s source problem with time-dependent fluid injection/extraction in an idealized poroelastic medium, also in a homogeneous domain. In a follow-up study, Bekele~\cite{bekele2024physics} further applied curriculum training to solve a poroelastic problem under idealized conditions. Millevoi et al.~\cite{millevoi4074416physics} addressed the 1D Terzaghi consolidation problem and the 2D Mandel problem, incorporating relatively strong coupling between poroelastic equations, but their approach relied heavily on substantial data extracted from sensors within the domain to support the training process. While these studies provide valuable insights into applying PINNs for poroelasticity, they remain focused on homogeneous systems, often dealing with weakly coupled equations or requiring additional training data to complement the physics-based learning. This highlights a significant research gap in applying PINNs to heterogeneous poroelastic media, especially in scenarios involving strong coupling and material discontinuities.

This paper seeks to bridge this gap by developing an efficient PINNs framework for modeling poroelasticity in heterogeneous media, effectively handling jumps and discontinuities across material interfaces, and managing strong equation coupling based purely on the underlying physics—without the need for additional training data. The specific objectives of this study are threefold:

\begin{itemize}
    \item To develop a novel framework of PINNs, called composite neural network (CoNN) to address coupled-field problems in poroelasticity with strong fluid-to-solid and solid-to-fluid coupling.
    \item To integrate the CoNN with the Interface-PINNs (I-PINNs) framework developed by Sarma et al.~\cite{sarma2024interface}, enabling the model to handle interfacial jumps in field variables, thereby making it suitable for multilayered media with material property discontinuities at interfaces.
    \item To benchmark the framework, by comparing it against established domain-decomposition-based frameworks available in literature like eXtended PINNs (XPINNs) framework~\cite{jagtap2020extended}, as well as a single neural network architecture.
\end{itemize}

The practical applications of this framework are significant, particularly as a numerical solver for interface problems. It holds promise for modeling heterogeneous subsurface and geothermal systems characterized by complex geological stratification and numerous material interfaces. The rest of the paper is organized as follows: Section~\ref{sec:goveq} outlines the theory and governing differential equations for poroelasticity in a heterogeneous domain. Section~\ref{sec:heterogeneous} provides an overview of PINNs, followed by a detailed description of the proposed PINNs architecture tailored for handling poroelasticity in heterogeneous materials. Here, the architecture of the proposed PINNs framework for heterogeneous poroelasticity is described, along with its implementation, and guidelines for improving accuracy and convergence are provided. In Section~\ref{sec:numerical}, the numerical results of the proposed PINNs architecture are compared with those obtained using the conventional PINNs approach. The evaluation encompasses numerical accuracy and computational costs on two specific 1D heterogeneous poroelastic problems. Following this analysis, a sensitivity assessment is conducted, wherein the performance of the proposed framework for each applied modification is evaluated. Section~\ref{sec:comparison_with_xpinns} presents the comparison of the proposed architecture with XPINNs for the set of problems outlined in Section~\ref{sec:numerical}. Section~\ref{sec:discussions} presents a discussion of the current study, its limitations, potential for improvements and provides an outlook for the work. Finally, Section~\ref{sec:conclusions} presents concluding remarks of this study.

\section{Governing Equations}
\label{sec:goveq}
\sloppy
The mechanical behavior of a porous medium is influenced by the presence of a moving fluid within it. Simultaneously, alterations in the mechanical state of the porous structure impact the fluid's conduct within the pores. The foundational principles of poroelasticity revolve around the interconnected phenomena of deformation and diffusion: fluid-to-solid coupling takes place when variations in fluid pressure or mass result in the deformation of the porous framework, while solid-to-fluid coupling occurs when changes in the stress of the porous structure lead to adjustments in fluid pressure or mass. In line with these phenomena, the time-dependent nature of the fluid-filled porous medium becomes apparent. Suppose the porous medium undergoes compression, leading to an increase in fluid pressure within the pores and subsequent fluid flow. The temporal evolution of fluid pressure, characterized by the dissipation of pressure through diffusive fluid flux according to Darcy's law, initiates a time-dependent modification in poroelastic stresses. These stresses, in turn, reciprocate by influencing the fluid pressure field. Evidently, the model depicting this process is time-dependent and can be considered quasi-static when inertial forces are neglected.

\begin{figure}[!hbt]
    \centering
    \includegraphics[width=0.55\textwidth]{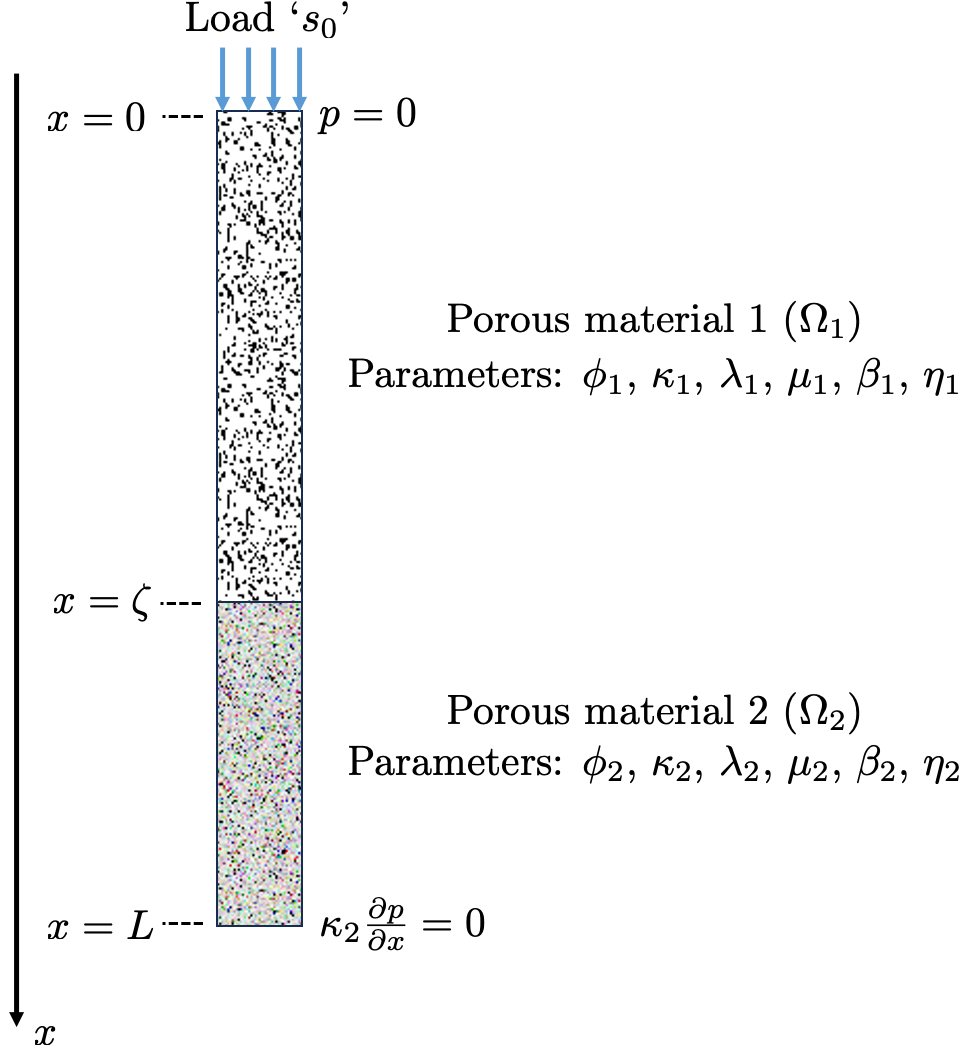}
    \caption{Problem domain in one dimension: The bottom surface is fixed, while the top surface is subjected to stress $s_0$; no-flux boundary condition is imposed on the bottom surface, while a zero-pressure condition is imposed on the top.}
    \label{problem_domain}
\end{figure}

The governing equations have been formulated under the assumptions of quasi-static, linearized poroelasticity \cite{coussy2004poromechanics} such that in a one-dimensional bounded domain $\Omega=[0,L]$, partitioned into two non-overlapping sub-domains $\Omega_{1}=[0,\zeta]$ and $\Omega_{2}=[\zeta,L]$ by a perfectly bonded material interface at $x=\zeta\in(0,L)$ such that $\Omega=\Omega_{1}\cup\Omega_{2}$. The displacements $u_{m}(x,t)$ and pressures $p_{m}(x,t)$ that satisfy the following equations simultaneously in each subdomain $\Omega_{m}$ (for $m=1,2,...,M$ where $M$ represents the total number of sub-domains) are sought:
\begin{gather}
    \frac{\partial}{\partial x}\left((\lambda_m+2\mu_m)\frac{\partial u_m}{\partial x}\right)+\alpha\frac{\partial p_m}{\partial x}=h_m(x,t), x\in\Omega_m, 0<t\leq T, \\
    \frac{\partial }{\partial t}\left(\phi_m\beta_mp_m+\frac{\partial u_m}{\partial x}\right)-\frac{\partial}{\partial x}\left(\frac{\kappa_m}{\eta_m}\frac{\partial p_m}{\partial x}\right)=q_m(x,t), x\in\Omega_m, 0<t\leq T,
\end{gather}
where, for each subdomain $\Omega_m$, $\lambda_m, \mu_m$ are the Lame’s parameters; $\alpha, \phi_m, \beta_m, \kappa_m$, and $\eta_m$ denote the Biot-Willis constant, porosity, fluid compressibility, intrinsic permeability, and fluid viscosity, respectively. Additionally, $q_m(x,t)$ is the volumetric source/sink term, and $h_m(x,t)$ is the external body force on the medium (both of which are taken to be zero). The quantities denoted by a subscript $m$ can be discontinuous across the interface. Figure~\ref{problem_domain} shows the schematic of the problem domain containing a poroelastic media with one interface. The boundary and initial conditions are given as:
\begin{subequations}
\begin{gather}
    (\lambda_1+2\mu_1)\frac{\partial u_1}{\partial x}=-s_0 \text{  at  } x=0 \text{  and  } u_2=0 \text{ at  } x=L,\\
    p_1=0 \text{ at  } x=0 \text{ and  } \frac{\partial p_2}{\partial x}=0 \text{  at  } x=L, \\
    \phi_m\beta_m p_m+\frac{\partial u_m}{\partial x}=0, \text{  for  }x\in\Omega, t=0 .
\end{gather}
\end{subequations}
The applied load on $x=0$ is given as $s_0$. The initial condition assumes a zero initial change in water content. To make the data suitable for deep learning frameworks, quantities were non-dimensionalized by employing the characteristic length $L$ and reference values of Lame's coefficients, permeability, and viscosity. This is achieved through the following variable substitution:
\[
\begin{aligned}[t]
    x &:= \frac{x}{L}, &
    t &:= \frac{(\lambda_0 + 2\mu_0)\kappa_0t}{\eta L^2}, &
    u_m &:= \frac{(\lambda_0 + 2\mu_0)u_m}{s_0L}, &
    p_m &:= \frac{p_m}{s_0}, &
    \nu_m &:= \frac{\lambda + 2\mu}{\lambda_0 + 2\mu_0},
\end{aligned}
\]
\[
\begin{aligned}[b]
    \kappa_m &:= \frac{\kappa_m/\eta_m}{\kappa_0/\eta_0}, &
    a_m &:= \phi_m\beta_m(\lambda_0+2\mu_0), &
    f_m(x,t) &:= \frac{\eta_0 L^2}{s_0\kappa_0}q_m(x,t).
\end{aligned}
\]

Here, the quantities with subscript `$0$' designates reference values for the associated parameters. Additionally, homogeneous boundary conditions are established using the substitution $u_m(x,t):=u_m(x,t)+x-1$. Consequently, the resulting nondimensionalized system becomes:
\begin{gather}
    -\frac{\partial}{\partial x}\left(\nu_m\frac{\partial u_m}{\partial x}\right)+\frac{\partial p_m}{\partial x}=0, x\in\Omega_m, 0<t\leq T, \label{lin_mom}\\
    \frac{\partial}{\partial t}\left(a_mp_m+\frac{\partial u_m}{\partial x}\right)-\frac{\partial}{\partial x}\left(\kappa_m\frac{\partial p_m}{\partial x}\right)=f_m(x,t), x\in\Omega_m, 0<t\leq T\label{mass_bal}, 
\end{gather}

with the boundary and initial conditions:
\begin{subequations}
\begin{gather}
    \nu_1\frac{\partial u_1}{\partial x}=0 \text{  at  } x=0 \text{  and  } u_2=0 \text{ at  } x=1,\label{eq:bc1}\\
    p_1=0 \text{ at  } x=0 \text{ and  }\kappa_2 \frac{\partial p_2}{\partial x}=0 \text{  at  } x=1,\label{eq:bc2} \\
    a_mp_m+\frac{\partial u_m}{\partial x}=1, \text{  for  }x\in\Omega, t = 0.\label{eq:ic}
\end{gather}
\end{subequations}
The interface conditions (at $x=\zeta$) ensure the continuity of fluxes and the primary variables at the interface such that
\begin{equation}
\begin{aligned}\label{continuity_cond}
    \llbracket u \rrbracket &= 0, & \Biggl\llbracket \nu \frac{\partial u}{\partial x} \Biggr\rrbracket  &= 0, & \llbracket p \rrbracket &= 0, & \Biggl\llbracket \kappa \frac{\partial p}{\partial x} \Biggr\rrbracket &= 0.
\end{aligned}
\end{equation}
The operator $\llbracket \bigodot \rrbracket=\bigodot_2-\bigodot_1$ represents a jump in the field $\bigodot$. The coefficients are piecewise constant across the interface, as follows:
\begin{equation}
\begin{aligned}\label{coeffs}
    \nu(x) &= \begin{cases}
                \nu_1, & x \leq \zeta \\
                \nu_2, & x > \zeta
              \end{cases}, &
    a(x) &= \begin{cases}
                a_1, & x \leq \zeta \\
                a_2, & x > \zeta
              \end{cases}, &
    \kappa(x) &= \begin{cases}
                \kappa_1, & x \leq \zeta \\
                \kappa_2, & x > \zeta
              \end{cases}.
\end{aligned}
\end{equation}

\section{Physics-informed Neural Networks for Heterogeneous Poroelastic Media}
\label{sec:heterogeneous}

\subsection{Physics-informed Neural Networks}
Over the last few decades, significant efforts have been devoted to leveraging ML approaches, such as support vector machines \cite{li2006support,ceryan2013modeling}, Gaussian processes \cite{joshi2023machine,williams2006gaussian} and artificial neural networks \cite{mujumdar2007artificial,hsiao2005fuzzy,al2012performance,yagawa1996neural} to model physical processes. Many of these approaches operate as data-driven mechanisms, treating ML as a black-box tool. However, the accuracy of these approximations is heavily reliant on the quality and quantity of the training data employed. This poses a substantial challenge in engineering applications, where issues like data accessibility, validity, noise, and various uncertainties come into play. Furthermore, there is no assurance that the ML models faithfully capture the underlying physics of the problem. A constructive solution to these challenges has emerged through the innovative application of PINNs. PINNs use deep neural networks (DNNs) to approximate solutions to the governing differential equations (ODEs/PDEs). DNNs are structured like the human brain, with interconnected hidden layers containing nodes called neurons. These models integrate the physics of the problem as soft constraints guiding the optimization process. The loss functional $\mathcal{L}(\theta)$ consists of the linear combination of the residual of PDEs, interface, boundary, and initial conditions. The goal is to minimize $\mathcal{L} (\theta)$ by optimizing the parameter set $\theta$ on randomly selected collocation points within the domain and on boundaries. 

Consider the time-varying PDE $\mathcal{P}u(x,t)=f(x,t)$, where $\mathcal{P}$ represents the differential operator, and $f(x,t)$ is the source function. The dependent variable $u(x,t)$ is defined on the domain $\Omega\times T$, with $\Omega\in \mathcal{R}^d$ and $T\in \mathcal{R}$ representing the spatial and temporal domains respectively. The real space and number of dimensions in the real space are denoted by $\mathcal{R}$ and $d$ respectively. The PDE is subjected to Dirichlet boundary condition $b_D(x,t)$ on $\Gamma_D$, Neumann boundary condition $b_N(x,t)$ on $\Gamma_N$, such that, $\Gamma_D\cup\Gamma_N=\partial\Omega$, where $\partial\Omega$ denotes the entire boundary of the computational domain $\Omega$. It is also subjected to the initial condition $u_0(x)$ at $t=0$. According to PINNs, the unknown variable $u(x,t)$ is approximated by a DNN $u(x,t)\approx\tilde{u}=\mathcal{N}(x,t,\theta)$. The flux at boundary $\Tilde{q}=\nabla\Tilde{u}\cdot\mathbf{n}$, as well as the PDE residual $\mathcal{P}u(x,t)-f(x,t)$ is computed using the automatic differentiation (AD) technique \cite{rall1981automatic}. The normal to the boundary is denoted by $\mathbf{n}$. Therefore, the total loss functional is defined as
\begin{equation}\label{pinn_loss}
\begin{aligned}
    \mathcal{L}&=\lambda_1||\mathcal{P}\tilde{u}(x,t)-f(x,t)||_{\Omega\times T} \\
               &+\lambda_2||\tilde{u}(x,t)-b_D||_{\Gamma_D\times T} \\
               &+\lambda_3||\tilde{q}(x,t)-b_N||_{\Gamma_N\times T} \\
               &+\lambda_4||\tilde{u}(x,t)-u_0||_{\Gamma_D\times t=0}
\end{aligned}.
\end{equation}
In Eq.~(\ref{pinn_loss}), the quantity $||\bigodot||$ represents an error metric, and the term $\lambda_i$ represents the penalty (weight) associated with $i$-th loss term. Two common error metrics are mean square errors (MSE) and sum of square errors (SSE). The loss functional is minimized over a set of collocation points sampled across the spatiotemporal domain $\Omega\times T$, the boundary $\partial\Omega\times T$, and at the initial condition $\Omega\times t=0$. 

As described before, the approximation $\tilde{u}(x,t)$ is obtained using a DNN, which is essentially a feed-forward network consisting of a collection of neurons arranged in consecutive layers. Moreover, the spatiotemporal inputs $(x,t)$ can be collectively wrapped into a single input vector $\mathbf{x}$ such that $\mathbf{x}=[x,t]$. Neurons in adjacent layers of the fully connected network are linked, whereas neurons within one single layer are not connected. For a deep neural network with $K$ layers consisting of an input layer, an output layer, and $K-2$ hidden layers, the network approximation can be represented as a combination of $K$ mathematical functions $\mathcal{A}_i (\mathbf{x}_{i},\theta_{i})$ as follows:
\begin{equation}\label{DNN}
    \Tilde{u}=\mathcal{N}(\mathbf{x},\theta)=\mathcal{A}_K\circ\mathcal{A}_{K-1}\circ\mathcal{A}_{K-2}\cdot\cdot\cdot\circ\mathcal{A}_1(\mathbf{x},\theta)
\end{equation}
The layer composition $\circ$ is to be read as $\mathcal{A}_{i+1}(\mathcal{A}_i(\mathbf{x},\theta))$. The function $\mathcal{A}$ can be a linear or non-linear function such that
\begin{equation}\label{activ}
    \mathcal{A}_i(\mathbf{x}_i,\theta_i)=\sigma(\mathbf{w}_i^T\mathbf{x}_i+\mathbf{b}_i)
\end{equation}
where, $\sigma$ is known as the activation function, $\mathbf{x}_i$ is the input vector to the $i$-th layer, and $\theta_i$ denotes the set of parameters for the $i$-th layer consisting of the weight matrix ($\mathbf{w}_i$) and bias vector ($\mathbf{b}_i$). A schematic of PINNs is shown in Figure~\ref{pinns_ipinns}(a).

\subsection{Interface PINNS (I-PINNs)}
\begin{figure}[!hbt]
    \centering
    \includegraphics[width=1.0\textwidth]{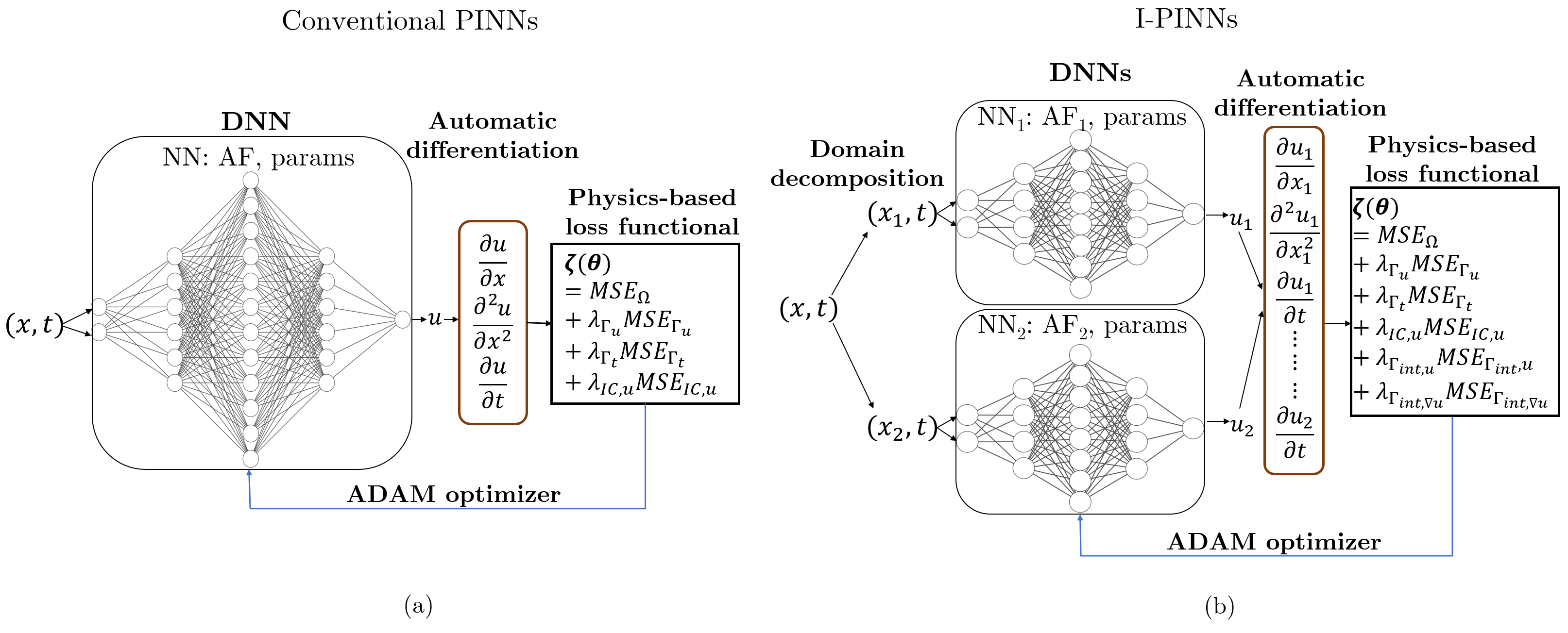}
    \caption{Schematic of (a) a conventional PINNs model, (b) I-PINNs for a problem in 1-D domain with two distinct regions divided by an interface. This modified setup involves the partitioning of input variables within these distinct domains, subsequently inputting them into separate neural networks utilizing different activation functions.}
    \label{pinns_ipinns}
\end{figure}

It is important to highlight that Eqs.~\eqref{mass_bal},~\eqref{eq:ic},~\eqref{continuity_cond}, and~\eqref{coeffs} may result in solutions with weak (slope) discontinuities. The conventional PINNs architecture, however, is not effective in handling discontinuous solution fields. As noted in previous studies \cite{selmic2002neural,sarma2023variational}, the traditional neural network architecture tends to produce uniform and continuous approximations throughout the domain, thereby failing to capture jumps (or kinks) at interfaces. Here, a modified PINNs framework termed Interface-PINNs (I-PINNs) is utilized. This framework was introduced previously by Sarma et al. \cite{sarma2024interface,sarma2023variational} due to its superior accuracy and convergence compared to conventional PINNs and other domain-decomposition-based PINNs (like M-PINN~\cite{zhang2022multi} and XPINNs~\cite{jagtap2020extended}) for interface problems. In I-PINNs, the domain is decomposed into homogeneous sub-domains, and separate neural networks are employed for each sub-domain. These networks share parameters while using distinct activation functions, as illustrated in Figure~\ref{pinns_ipinns}(b). Interface conditions, as described by Eq.~\eqref{continuity_cond}, are incorporated as additional terms in the loss functional. Notably, the neural networks across the interface communicate through a shared set of parameters and interface conditions. Simultaneously, the use of different activation functions on either side of the interface allows for flexibility in approximating discontinuities at the interface. An improved version of I-PINNs, termed Adaptive I-PINNs (AdaI-PINNs), was developed by Roy et al.~\cite{roy2024adaptive}. This version further enhances the computational efficiency of I-PINNs, particularly for materials with multiple interfaces. However, in this study, we have restricted our usage to the original I-PINNs framework, as the numerical examples outlined here involve materials with only one interface.

\subsection{I-PINNs for Coupled PDEs}
\begin{figure}[!hbt]
    \centering
    \includegraphics[width=1.0\textwidth]{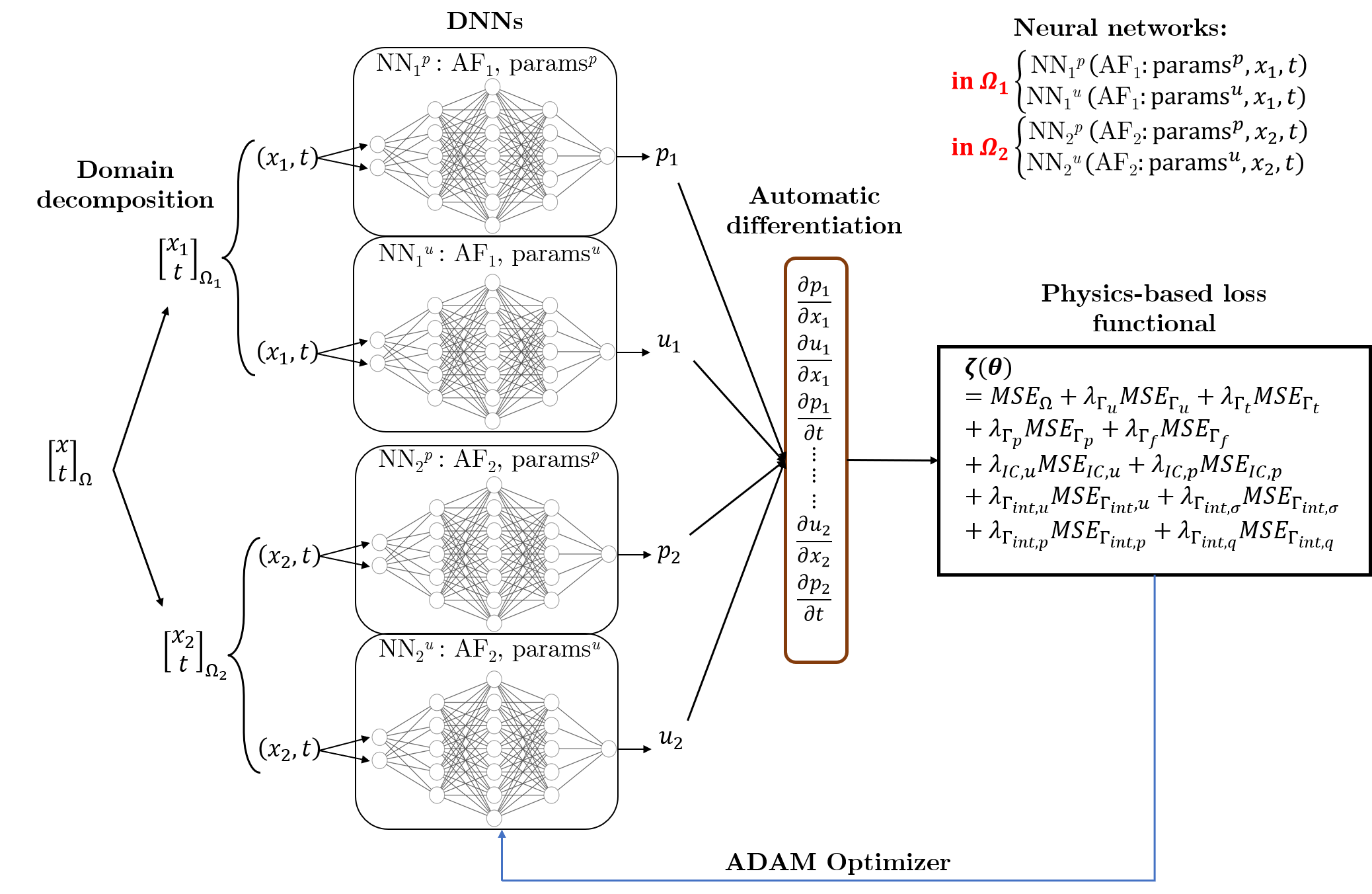}
    \caption{I-PINNs for 1-D coupled poroelasticity with an interface using a composite neural network (CoNN) architecture. Networks $\text{NN}_1^p$ and $\text{NN}_2^p$ model pressures across $\Omega_1$ and $\Omega_2$ with identical parameters ($\text{params}^p$) but distinct activation functions ($\text{AF}_1$ and $\text{AF}_2$), while $\text{NN}_1^u$ and $\text{NN}_2^u$ model displacements across $\Omega_1$ and $\Omega_2$ with shared parameters ($\text{params}^u$) and different activation functions ($\text{AF}_1$ and $\text{AF}_2$). In a nutshell: across subdomains, activation functions vary; between fields, only parameters differ.}
    \label{cnn_schematic}
\end{figure}

As described in Section~\ref{sec:goveq}, the equations of poroelasticity in heterogeneous media require the solution of displacements, $u_m$, and pressures, $p_m$, in each subdomain, $\Omega_m$. As such the I-PINNS framework must be appropriately modified for such coupled-field problems. A composite neural network (CoNN) framework is proposed, featuring distinct neural networks for each output field variable (i.e., displacements and pressures) within each subdomain. These neural networks have the same activation functions but different sets of parameters. However, to address discontinuities at material interfaces, this architecture is combined with the I-PINNs framework. In this combined approach, neural networks for distinct sub-domains share the same set of parameters but utilize different activation functions. As such, every output field variable gets a neural network with the same set of parameters throughout the problem domain, but with differing sets of activation functions across any material interface. However, the parameter set used in the neural networks is different for every output field variable. Figure~\ref{cnn_schematic} presents a schematic of the proposed architecture.

\begin{figure}[!hbt]
    \centering
    \includegraphics[width=0.75\textwidth]{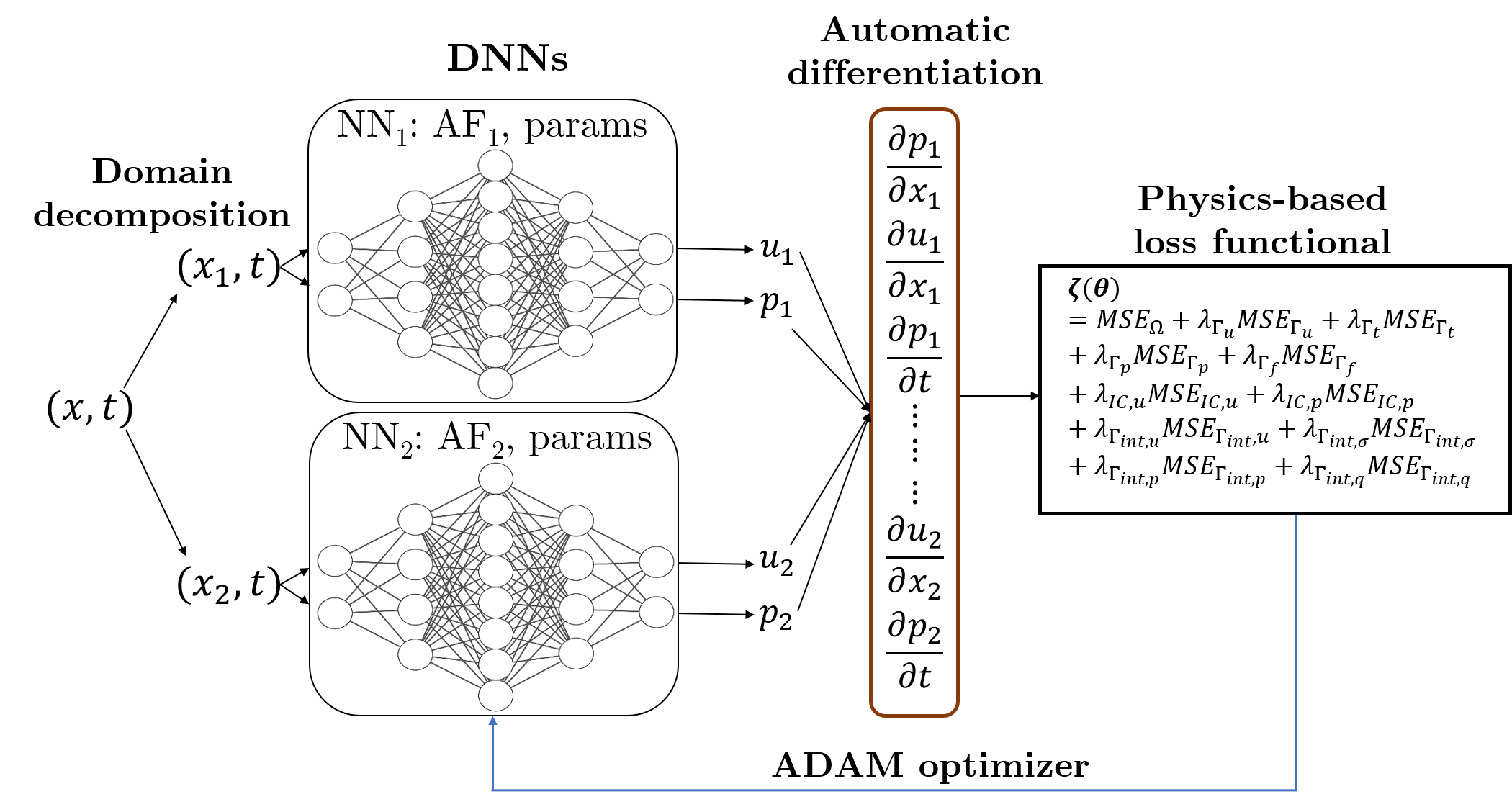}
    \caption{I-PINNs architecture for 1-D coupled poroelasticity with an interface using a single neural network (SNN) for each subdomain ($\text{NN}_1$ and $\text{NN}_2$ for $\Omega_1$ and $\Omega_2$ respectively). Each neural network outputs two field variables (pressures and displacements) for their respective sub-domains. These neural networks share the same set of parameters ‘params’ but use different activation functions $\text{AF}_1$ and $\text{AF}_2$ (in accordance to I-PINNs).}
    \label{snn_schematic}
\end{figure}

The architecture proposed in Figure~\ref{cnn_schematic} is to be contrasted with a straightforward extension of the I-PINNs architecture to poroelasticity problems in heterogeneous media where a single neural network (SNN) would feature two output neurons within each subdomain, each dedicated to representing a specific output field. While these neural networks would share the same set of parameters, they would employ different activation functions across a material interface, as in Figure~\ref{snn_schematic}.

\begin{figure}[!hbt]
    \centering
    \includegraphics[width=0.8\textwidth]{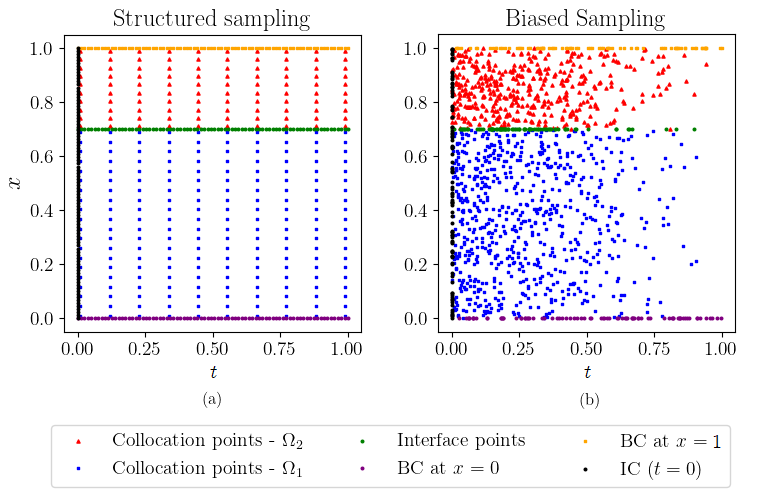}
    \caption{Representative figure where training points are sampled from – inside the two sub-domains (collocation points in $\Omega_1$ and $\Omega_2$), on the two boundaries (BC at $x=0$ and $x=1$) and at the initial condition (IC at $t=0$), using two sampling techniques: (a) in a structured grid, (b) in a biased manner.}
    \label{sampling}
\end{figure}

Regardless of the type of architecture used (SNN or CoNN), the total loss functional $\mathcal{L}(\theta)$ aggregates mean square errors (MSEs) from various terms, given as:
\begin{multline}
    \mathcal{L}(\theta)=\lambda_{\Omega}\text{MSE}_{\Omega}+\lambda_{\Gamma_{u}}\text{MSE}_{\Gamma_{u}}+\lambda_{\Gamma_{t}}\text{MSE}_{\Gamma_{t}}+\lambda_{\Gamma_{p}}\text{MSE}_{\Gamma_{p}}+\lambda_{\Gamma_{f}}\text{MSE}_{\Gamma_{f}}+\lambda_{IC,u}\text{MSE}_{IC,u}\\+\lambda_{IC,p}\text{MSE}_{IC,p}+\lambda_{\Gamma_{int,u}}\text{MSE}_{\Gamma_{int,u}}+\lambda_{\Gamma_{int,\sigma}}\text{MSE}_{\Gamma_{int,\sigma}}+\lambda_{\Gamma_{int,p}}\text{MSE}_{\Gamma_{int,p}}+\\\lambda_{\Gamma_{int,q}}\text{MSE}_{\Gamma_{int,q}}  
\end{multline}
The various loss terms in the above functional, include governing PDEs (linear momentum and mass balance equations summed together as $\text{MSE}_{\Omega}$), Dirichlet and Neumann boundary conditions for linear momentum balance ($\text{MSE}_{\Gamma_{u}}$ and $\text{MSE}_{\Gamma_{t}}$), and mass balance ($\text{MSE}_{\Gamma_{p}}$, and $\text{MSE}_{\Gamma_{f}}$), the initial conditions for displacements ($\text{MSE}_{IC,u}$), and pressures ($\text{MSE}_{IC,p}$), and the interface compatibility conditions for displacements, tractions, pressures, and fluxes ($\text{MSE}_{\Gamma_{int,u}}$, $\text{MSE}_{\Gamma_{int,\sigma}}$, $\text{MSE}_{\Gamma_{int,p}}$, and $\text{MSE}_{\Gamma_{int,q}}$ respectively). These are further represented as:
\begin{subequations}
    \begin{equation}
        \begin{gathered}
            \text{MSE}_{\Omega} = \frac{1}{N_{\Omega\times T}} \left(\begin{aligned}
                &\sum_{i=1}^{N_{\Omega\times T}}\left[-\frac{\partial}{\partial x}\left(\nu\frac{\partial \tilde u}{\partial x}\right) + \frac{\partial\tilde p}{\partial x}\right] \\
                &+ \sum_{i=1}^{N_{\Omega\times T}}\left[\frac{\partial}{\partial t}\left(a\tilde p+\frac{\partial\tilde u}{\partial x}\right) - \frac{\partial}{\partial x}\left(\kappa\frac{\partial\tilde p}{\partial x}\right) - f_m(x,t)\right]
            \end{aligned}\right)\\
        \end{gathered}
    \end{equation}
    \begin{equation}
        \begin{gathered}
            \text{MSE}_{\Gamma_{u}}=\frac{1}{N_{\Gamma_{u}}}\left(\sum_{i=1}^{N_{\Gamma_{u}}}{\left[\tilde u-\overset{\circ}{u_D}\right]}\right)\\
        \end{gathered}
    \end{equation}
    \begin{equation}
        \begin{gathered}
            \text{MSE}_{\Gamma_{t}}=\frac{1}{N_{\Gamma_{t}}}\left(\sum_{i=1}^{N_{\Gamma_{t}}}{\left[\nu\frac{\partial \tilde u}{\partial x}-\overset{\circ}{u_N}\right]}\right)\\
        \end{gathered}
    \end{equation}
    \begin{equation}
        \begin{gathered}
            \text{MSE}_{\Gamma_{p}}=\frac{1}{N_{\Gamma_{p}}}\left(\sum_{i=1}^{N_{\Gamma_{p}}}{\left[\tilde p-\overset{\circ}{p_D}\right]}\right)\\
        \end{gathered}
    \end{equation}
    \begin{equation}
        \begin{gathered}
            \text{MSE}_{\Gamma_{f}}=\frac{1}{N_{\Gamma_{f}}}\left(\sum_{i=1}^{N_{\Gamma_{f}}}{\left[\kappa\frac{\partial \tilde p}{\partial x}-\overset{\circ}{p_N}\right]}\right)\\
        \end{gathered}
    \end{equation}
    \begin{equation}
        \begin{gathered}
            \text{MSE}_{\Gamma_{IC}}=\frac{1}{N_{\Gamma_{IC}}}\left(\sum_{i=1}^{N_{\Gamma_{IC}}}{\left[\tilde u-\overset{\circ}{u_{IC}}\right]}\right)\\
        \end{gathered}
    \end{equation}
    \begin{equation}
        \begin{gathered}
            \text{MSE}_{\Gamma_{IC}}=\frac{1}{N_{\Gamma_{IC}}}\left(\sum_{i=1}^{N_{\Gamma_{IC}}}{\left[\tilde p-\overset{\circ}{p_{IC}}\right]}\right)\\
        \end{gathered}
    \end{equation}
    \begin{equation}
        \begin{gathered}
            \text{MSE}_{\Gamma_{int}}=\frac{1}{N_{\Gamma_{int}}}\left(\sum_{i=1}^{N_{\Gamma_{int}}}{\llbracket\tilde u\rrbracket}\right)\\
        \end{gathered}
    \end{equation}
    \begin{equation}
        \begin{gathered}
            \text{MSE}_{\Gamma_{int}}=\frac{1}{N_{\Gamma_{int}}}\left(\sum_{i=1}^{N_{\Gamma_{int}}}{\Biggl\llbracket\nu\frac{\partial\tilde u}{\partial x}\Biggr\rrbracket}\right)\\
        \end{gathered}
    \end{equation}
    \begin{equation}
        \begin{gathered}
            \text{MSE}_{\Gamma_{int}}=\frac{1}
            {N_{\Gamma_{int}}}\left(\sum_{i=1}^{N_{\Gamma_{int}}}{\llbracket\tilde p\rrbracket}\right)\\[-0.5em]
        \end{gathered}
    \end{equation}
    \begin{equation}
        \begin{gathered}
            \text{MSE}_{\Gamma_{int}}=\frac{1}{N_{\Gamma_{int}}}\left(\sum_{i=1}^{N_{\Gamma_{int}}}{\Biggl\llbracket\kappa\frac{\partial\tilde p}{\partial x}\Biggr\rrbracket}\right)
        \end{gathered}
    \end{equation}
\end{subequations}\label{eq:loss_terms}
$N_{\Omega\times T}$ is the number of collocation points in the spatiotemporal domain over which the residual of the PDE is minimized. $N_{\Gamma_u}$ and $N_{\Gamma_t}$ represent the number of training points on boundaries, specifying Dirichlet and Neumann boundary conditions for displacements respectively, while $N_{\Gamma_p}$ and $N_{\Gamma_q}$ denote the corresponding number of training points for pressures on the boundaries. $N_{\Gamma_{IC}}$ and $N_{\Gamma_{int}}$ denote the number of training points on the initial condition and interface respectively. $\lambda_i$ represents the corresponding relative weight (penalty) for each loss term $\text{MSE}_i$ with respect to the weight for the loss term of the governing PDEs. Also, $\overset{\circ}{\odot}$ implies the boundary or initial values for each corresponding term. The DNN approximates the output fields $\tilde u$ and $\tilde p$. Points in the spatiotemporal domain are extracted through biased sampling, where the density decreases over time, as illustrated in Figure~\ref{sampling}(b). The biased sampling approach is to be contrasted with structured sampling shown in Figure~\ref{sampling}(a). The biased sampling approach prioritizes early-time solutions in accordance with causality and therefore, yields significantly improved results compared to the structured sampling approach \cite{guo2022novel}. Additionally, points were randomly sampled while retaining the bias, in contrast to the structured grid sampling. For brevity, only the results obtained from the neural networks utilizing training points obtained through a biased sampling approach are presented.

The optimal set of parameters for all experiments outlined in this paper is obtained using the Adam optimization algorithm \cite{kingma2014adam} with an initial learning rate of $10^{-3}$. Training is conducted using the functions provided by the JAX library \cite{jax2018github}, a robust numerical computing library designed for high-performance machine learning research. The computations are executed on the NVIDIA T4 GPU, which is accessible through the Google Colaboratory platform \cite{bisong2019google}.

\subsection{Glorot Initialization and Hard Enforcement of Boundary/initial Conditions}
In recent years, various strategies have been explored to enhance the performance of PINNs in solving PDEs \cite{chiu2022can,peng2022rpinns,wang2023expert}. These efforts often involve designing more effective neural network architectures or improving training algorithms, incorporating techniques such as proper non-dimensionalization of equations, efficient sampling methods, the use of adaptive activation functions, and positional embeddings, among others. In this study, the two benchmark problems are initially solved using the proposed CoNN framework in its vanilla form. Subsequently, two special modifications are applied: Glorot initialization and hard enforcement of boundary and initial conditions (B/ICs). Detailed descriptions of these modifications are provided in the subsections that follow.

\subsubsection{Glorot initialization}
Before the commencement of training, the parameters of our DNN undergo initialization through a normally distributed random variable, with a mean of zero and a standard deviation specified by a scaling factor (set to 0.1). This random initialization is applied to both the weight matrices and bias vectors of each layer in the neural network. However, this initialization is found to play a crucial role in the effectiveness of gradient descent algorithms. The Glorot initialization scheme \cite{glorot2010understanding} is followed to improve the performance of the method. The scheme involves initializing the weights of a neural network layer by drawing random values from a distribution with a mean of $0$ and a variance calculated as $\frac{2}{n_{\text{in}}+n_{\text{out}}}$, where $n_{\text{in}}$ is the number of input units, and $n_{\text{out}}$ is the number of output units in the layer. The biases are set to zero. The Glorot initialization helps prevent issues such as vanishing or exploding gradients during training by appropriately scaling the weights, thereby promoting stable and efficient convergence in deep learning models.

\subsubsection{Hard enforcement of boundary/initial conditions}
In the context of PINNs, the conventional approach to applying B/ICs is to penalize the discrepancy of initial and boundary constraints for PDEs in a ``soft manner." Alternatively, a more robust strategy is to enforce these conditions in a ``hard" manner by incorporating specific solutions that rigorously satisfy the B/ICs \cite{deng2023physical,berrone2022enforcing,liu2022unified}. This augmentation significantly enhances the neural networks' capability to grapple with intricate geometric challenges. 

When the initial condition (IC) is ``soft-enforced," the deep neural network $\mathcal{N}(\mathbf{x},\theta)$ approximates the function $u(\mathbf{x})$ as $u(\mathbf{x})\approx\tilde{u}(\mathbf{x})=\mathcal{N}(\mathbf{x}\theta)$ with the initial condition of $u_0(\mathbf{x})$ at $t=0$. Conversely, in the case of ``hard enforcement," a new function $\tilde{u}$ is created to approximate $u(x,t)$ while satisfying the initial condition for all $\mathcal{N}(\mathbf{x},\theta)$, given by:
\begin{equation*}
    \tilde{u}(\mathbf{x})=u_0(\mathbf{x})+t\mathcal{N}(\mathbf{x},\theta).
\end{equation*}

For the ``hard-enforcement" of the Dirichlet boundary condition, where $u_{D1}$ is imposed at $x=x_{\text{min}}$ and $u_{D2}$ at $x=x_{\text{max}}$, a function $\tilde{u}$ is created to approximate $u(\mathbf{x})$ as follows, 
\begin{equation*}
    \tilde{u}(\mathbf{x})=(x-x_{\text{min}})(x-x_{\text{max}})\mathcal{N}(\mathbf{x},\theta) - u_{D1}\frac{x-x_{\text{max}}}{x_{\text{max}}-x_{\text{min}}} + u_{D2}\frac{x-x_{\text{min}}}{x_{\text{max}}-x_{\text{min}}}.
\end{equation*}

In both scenarios, there is no explicit need to specify the B/ICs in the loss functional, as the function $\tilde{u}(\mathbf{x})$ automatically satisfies the corresponding B/ICs.

\section{Numerical Experiments}
\label{sec:numerical}

In this section, the performance of the two proposed frameworks, namely, I-PINNs with CoNN and I-PINNs with SNN, is compared with each other, as well as with conventional PINNs. This comparison is conducted on two benchmark examples examining poroelasticity in heterogeneous materials with a single interface. In Section ~\ref{sec:improvement}, a comparative analysis is presented by solving the same benchmark problems with and without Glorot initialization and hard enforcement of B/ICs. To rigorously evaluate the accuracy of the proposed framework in addressing poroelastic problems, it's imperative to compare the approximations with closed-form analytical solutions. Although such solutions are scarce for heterogeneous poroelastic scenarios, a few have been found for 1D problems \cite{bean2014immersed}. The results of this study were directly compared with the exact solutions provided in this literature. All pertinent details, including the problem domain, interface location, material properties, and the closed-form analytical solutions for both problems, are extracted from Bean and Yi \cite{bean2014immersed}.

It is important to note that while the exact solutions conform to the boundary conditions (Eqs. \eqref{eq:bc1} and \eqref{eq:bc2}), they no longer satisfy the initial condition (Eq. \eqref{eq:ic}). The initial condition is determined by evaluating the given exact solutions at $t=0$. For conciseness, the collectively proposed methods are referred to as I-PINNs (CoNN and SNN), while the conventional PINNs method is denoted simply as PINNs. The comparisons are based on two key factors: computational cost i.e. run-time for training the models and accuracy. The root mean square error (RMSE) metric is used to compare the numerical accuracy of the experiments, while for quantifying the relative computational cost of an experiment, we define the metric cost as: $C = t_m/t_P$, where $t_m$  represents the total training time for the method under consideration, and $t_P$ represents the total training time for the proposed method. 

Furthermore, both frameworks are enhanced with various modifications, as described in the previous section, and the impact of each modification on the accuracy of approximations is assessed. Finally, the convergence profiles of the CoNN and SNN models for the test cases are investigated.

\subsection{Hyperparameter-tuning:}\label{subsec:hyperparametertuning}

Firstly, certain standards, such as the neural network architecture, total number of iterations, parameter scale for initialization, etc., must be established to ensure consistent comparisons. To determine these standards, the hyper-parameters are tuned on a simple 1D homogeneous problem involving a compressible fluid. The material constants and the analytical solution for this problem are provided below -

\begin{equation}\label{homo_parameters}
\begin{aligned}
    \nu &= 1, & \kappa &= 1, & a &= 0.01.
\end{aligned}
\end{equation}
\begin{equation}\label{homo_anal}
\begin{aligned}
    p(x,t) &=\cos\left(\frac{20}{3}\right)\sin\left(x\right)e^{-100t/101}, \\
    u(x,t) &= -\cos\left(\frac{20}{3}\right)\cos\left(x\right)e^{-100t/101}.
\end{aligned}
\end{equation}

It is noteworthy that for this specific problem, the revised boundary conditions are derived from Eq. \eqref{homo_anal}, given that the original boundary conditions (Eqs. \eqref{eq:bc1} and \eqref{eq:bc2}) are no longer applicable. In the CoNN framework, the hyperbolic-tangent (tanh) activation function is employed for training over 70,000 iterations with an initial learning rate of $10^{-3}$, initializing parameters at a scale of 0.1. The training set includes 5200 points, comprising 4900 collocation points, 100 points on each boundary, and 100 points at the initial condition. The RMSE for each potential architecture is computed, considering combinations of 1, 2, 3, 4 hidden layers and 5, 10, 20, 40, 80 hidden neurons in each layer, resulting in a total of 20 possible architectures. The weight (penalty) on each loss term is kept as 1.

\begin{figure}[!hbt]
    \centering
    \includegraphics[width=0.9\textwidth]{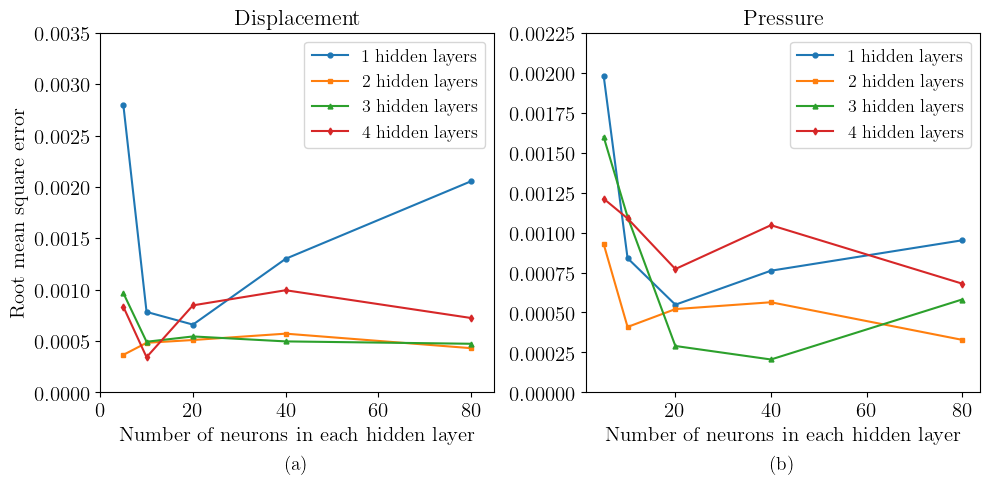}
    \caption{The plots illustrate the root mean square errors (RMSEs) corresponding to the approximations of displacements and pressure fields by our proposed I-PINNs with CoNN framework for the homogeneous compressible fluid problem. Various combinations of hidden layers (1,2,3,4) and neurons per layer (5,10,20,40,80) are examined, providing insights into the sensitivity of the model to different architectural configurations.}
    \label{homo_archi}
\end{figure}

\begin{figure}[!hbt]
    \centering
    \includegraphics[width=1.0\textwidth]{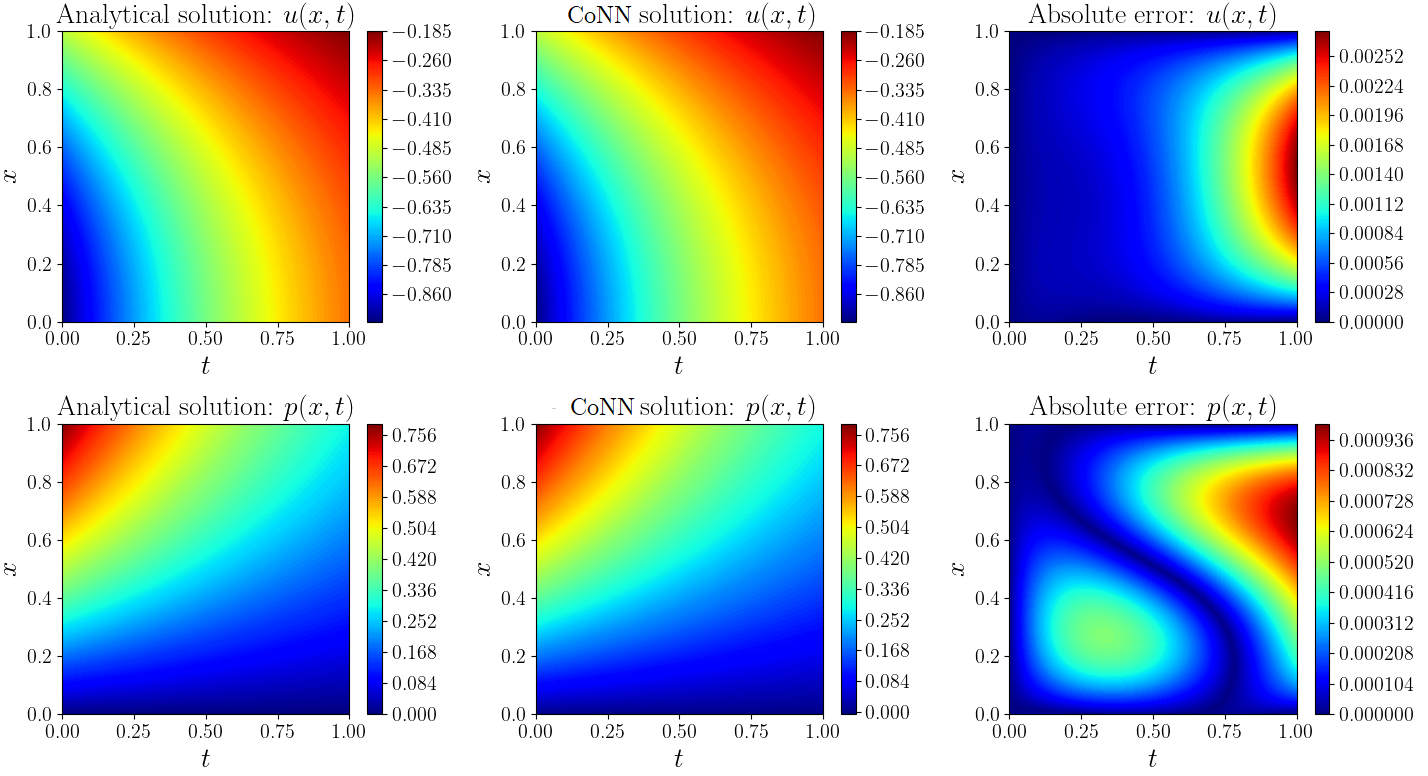}
    \caption{The displacements and pressure fields, accompanied by their respective absolute error contour plots, as approximated by the proposed I-PINNs with CoNN architecture for a homogeneous compressible fluid flow-problem. The activation functions GELU and tanh are employed across the two subdomains separated by the interface.}
    \label{homo_results}
\end{figure}

From Figure~\ref{homo_archi}, it is evident that with a single layer and 5 hidden neurons, the RMSE is high for both displacement and pressure. As the number of neurons increases to 20, the root mean square rrror (RMSE) decreases, but further increments result in an upsurge. Analyzing the displacement plot reveals that this issue is promptly addressed by incorporating multiple hidden layers. With multiple hidden layers, even using only 5 hidden neurons in each layer results in low RMSEs. However, for pressures, the task is more challenging, necessitating a higher number of neurons in each layer and more hidden layers to reduce the RMSE. Given that the aim is to deal with coupled poroelasticity problems, it is very important to choose a specific layer architecture that consistently results in relatively low errors in approximating both output fields – displacements and pressures. In this regard, while a 4 hidden layer architecture with 10 neurons each may yield a lower RMSE for displacements (as seen in Figure~\ref{homo_archi}(a)), it performs poorly for the pressure field (indicated in Figure~\ref{homo_archi}(b)). Therefore, it is essential to choose a layer architecture that minimizes the RMSE for both output fields, rather than optimizing for just one. Among the options, two architectures meet this criterion: 3 layers with 40 neurons each and 2 layers with 80 neurons each. The 3-layer configuration has a total of 6,804 parameters, whereas the 2-layer configuration has 13,602 parameters and is more computationally intensive. For practical purposes, the 3-layer architecture with 40 neurons each is selected for all subsequent experiments due to its simplicity and lower computational cost.

The approximated pressure and displacement plots, along with their corresponding absolute error contour plots, are presented in Figure~\ref{homo_results}. For this simulation, we employed the aforementioned layer architecture consisting of 3 layers with 40 neurons each. It is evident that this specific layer architecture has effectively approximated both fields, with errors mostly lying close to $t=1$ (near the characteristic time).

The discussion now transitions to two specific 1D heterogeneous problems, one featuring an incompressible fluid and the other involving a compressible fluid. The details for each are provided in the subsequent subsections.

\subsection{Problem 1: Heterogeneous Poroelastic Media with an Incompressible Fluid}\label{subsec:incompressible_fluid_numerical}

First, the problem described in Eqs. \eqref{lin_mom}-\eqref{mass_bal} is considered without any volumetric source/sink terms (i.e., $f_1(x,t)=f_2(x,t)=0$) and with incompressible fluid in both subdomains (i.e., $a_1=a_2=0$). The computational domain is considered as $x\in(0,1), 0\leq t\leq 1$, with a fixed material interface at $x=\zeta=1/5$, that divides the problem domain into two non-overlapping subdomains, $x_1\in(0,\zeta)$, and $x_2\in(\zeta,1)$. The elastic constants and the conductivities are specified as
\begin{equation}\label{prob1_constants}
\begin{aligned}
    \nu(x) &= \begin{cases}
                \nu_1=1, & x \leq \zeta \\
                \nu_2=\frac{\tan(1/15)\tan(12\pi/5)}{9\pi}, & x > \zeta
              \end{cases},\\
    a(x) &= \begin{cases}
                a_1=0, & x \leq \zeta \\
                a_2=0, & x > \zeta
              \end{cases},\\
    \kappa(x) &= \begin{cases}
                \kappa_1=1, & x \leq \zeta \\
                \kappa_2=\frac{1}{\tan(1/15)tan(12\pi/5)}, & x > \zeta
              \end{cases}.
\end{aligned}
\end{equation}
The displacements, tractions, pressures, and fluxes are assumed to be continuous at the interface as per Eq.~\eqref{continuity_cond}. The analytical solutions for pressures and displacements for the problem are given below
\begin{equation}\label{prob1_analsol}
\begin{aligned}
    p(x,t) &= \begin{cases}
                \cos\left(\frac{12\pi}{5}\right)\sin\left(\frac{x}{3}\right)e^{-t/9}, & x\leq\zeta \\
                \sin\left(\frac{1}{15}\right)\cos(3\pi(1-x))e^{-t/9}, & x>\zeta
              \end{cases},\\
    u(x,t) &= \begin{cases}
                -3\cos\left(\frac{12\pi}{5}\right)\cos\left(\frac{x}{3}\right)e^{-t/9}, & x \leq \zeta \\
                -\frac{3\cos(1/15)}{\tan(12\pi/5)}3\sin(3\pi(1-x))e^{-t/9}, & x > \zeta
              \end{cases},
\end{aligned}
\end{equation}

Approximations to displacements and pressures in both sub-domains were constructed using both PINNs and I-PINNs (CoNN and SNN) for this model problem using the hyperparameters detailed in Table~\ref{prob1_table}. It was observed that when using identical hyperparameters, I-PINNs with CoNN achieved an RMSE of $\mathcal{O}(10^{-4})$ for both displacements and pressures, while I-PINNs with SNN achieved an RMSE of $\mathcal{O}(10^{-3})$, which is one order lower. However, conventional PINNs were only able to obtain RMSE of $\mathcal{O}(10^{-2})$. It is also interesting to observe that although I-PINNs with CoNN use almost twice the number of parameters than that of I-PINNs with SNN, the computation time taken by the latter is marginally higher (by 9\%) than the former.

\begin{table}[!hbt]
\centering
\caption{Hyper-parameters used in training both PINNs \& I-PINNs (CoNN and SNN) models for approximating the solution to the incompressible fluid problem. The activation functions are hyperbolic-tangent (tanh) and Gaussian error linear unit (GELU).}
\label{prob1_table}
{\footnotesize
\begin{tabular}{p{4cm}|c|c|c}

    \hline
   \textbf{Parameter} & \textbf{I-PINNs} & \textbf{I-PINNs} & \textbf{PINNs} \\
   & \textbf{(CoNN)} & \textbf{(SNN)} &  \\
   \hline
   Layer architecture & [2,40,40,40,1]$\times$2 & [2,40,40,40,2] & [2,40,40,40,1]$\times$2 \\
   \hline
   Activation functions & GELU-tanh & GELU-tanh & GELU \\
   \hline
   Number of training points [domain interior, external boundaries, interface] & [4900,140,70] & [4900,200,100] & [4900,200,100] \\
   \hline
   Penalty [interface conditions, boundary conditions, initial condition] & [200, 500, 300] & [200, 500, 300] & [200, 500, 300] \\
   \hline
   Maximum iterations & 70000 & 70000 & 70000 \\
   \hline
   RMSE (displacements) & 2.6$\times 10^{-4}$  & 3.7$\times 10^{-3}$ & 5.4$\times 10^{-2}$ \\
   \hline
   RMSE (pressures) & 4.2$\times 10^{-4}$ & 4.8$\times 10^{-3}$ &  4.0$\times 10^{-2}$ \\
   \hline
   Cost & 1 & 1.2 & 0.8 \\
   \hline
\end{tabular}
}
\end{table}

\begin{figure}[!hbt]
    \centering
    \includegraphics[width=1.0\textwidth]{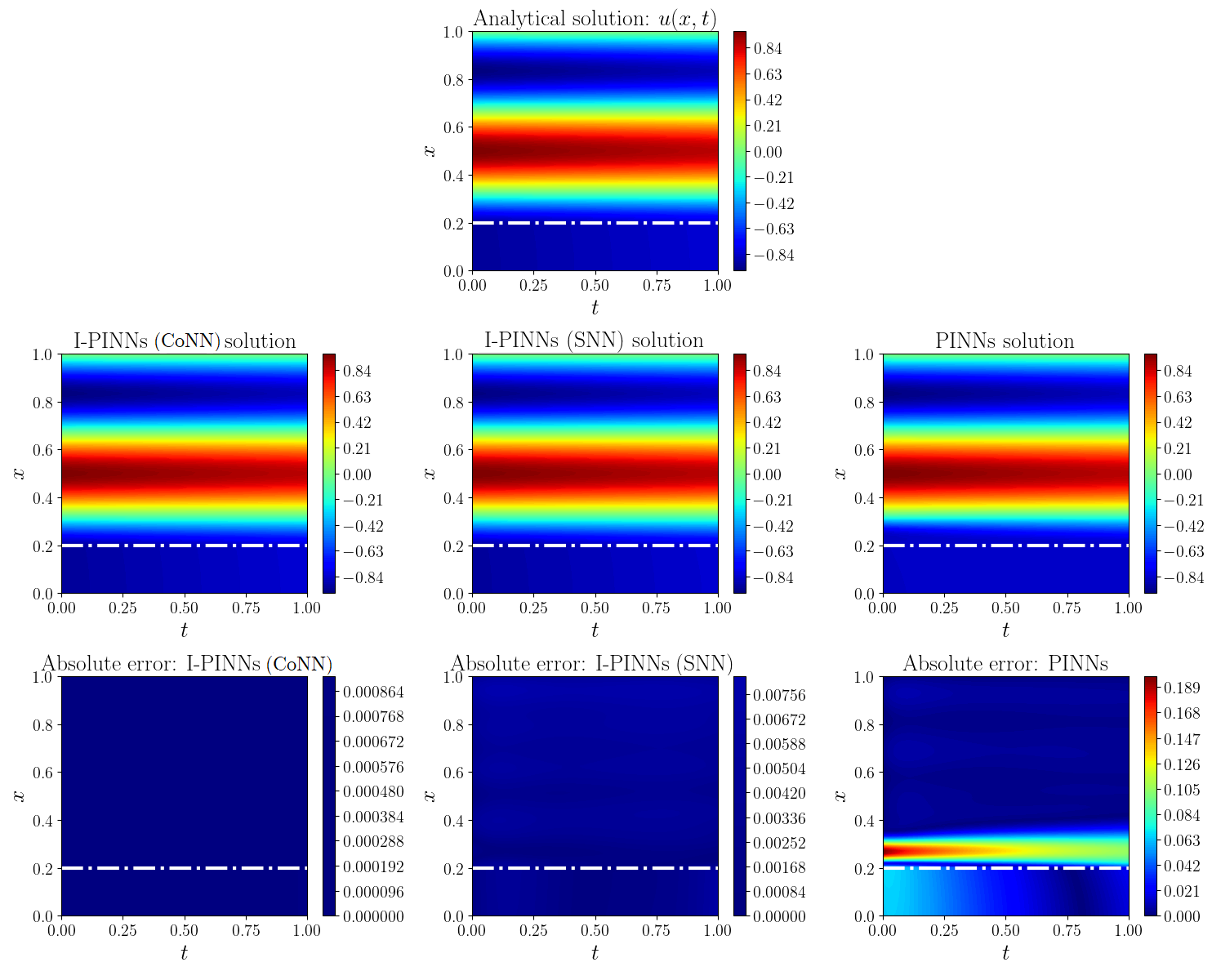}
    \caption{Displacement fields obtained by the two I-PINNs frameworks – SNN and CoNN, and as well as conventional PINNs (second row) for the incompressible fluid problem ($a=0$). The third row displays the corresponding absolute errors. The interface (at $x=\zeta=1/5$) is demarcated with a white dash-dot line.}
    \label{prob1_u}
\end{figure}

\begin{figure}[!hbt]
    \centering
    \includegraphics[width=1.0\textwidth]{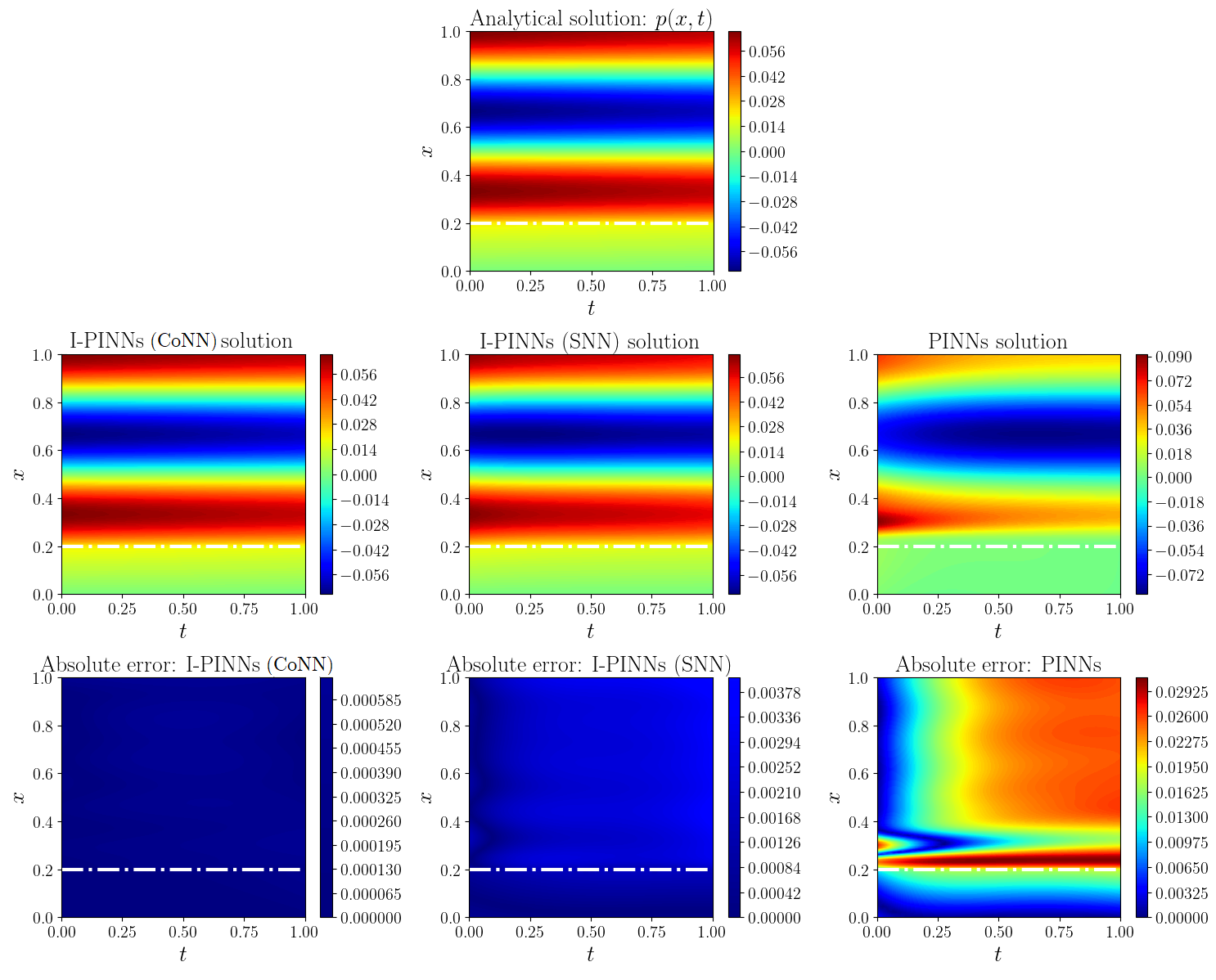}
    \caption{Pressure fields obtained by the two I-PINNs frameworks – SNN and CoNN, and as well as conventional PINNs (second row) for the incompressible fluid problem ($a=0$). The third row displays the corresponding absolute errors. The interface (at $x=\zeta=1/5$) is demarcated with a white dash-dot line.}
    \label{prob1_p}
\end{figure}

The results (for both displacements and pressures) and the corresponding absolute errors obtained from both the I-PINNs frameworks (SNN and CoNN) as well as PINNs are shown in Figs.~\ref{prob1_u} and~\ref{prob1_p}, respectively. Note that the maximum error occurs near the interface for PINNs. This is expected since the material property mismatch results in discontinuous gradients for pressures and displacements at the interface, and PINNs struggle to approximate these weak discontinuities accurately. Among the two I-PINNs models, the CoNN model approximated the fields with lesser errors compared to that of SNN. The maximum error in SNN is at least twice that of CoNN for both the displacement and pressure fields.

\subsection{Problem 2: Heterogeneous Poroelastic Media with a Compressible Fluid}\label{subsec:compressible_fluid_numerical}

As a second test problem, the problem described in Section 4.1 is modified by relaxing the assumption of fluid incompressibility. The problem domain remains identical, but the interface is now at $x=\zeta=1/3$. The new material coefficients and the subsequent analytical solutions for pressures and displacements are listed below:

\begin{equation}\label{prob2_constants}
\begin{aligned}
    \nu(x) &= \begin{cases}
                \nu_1=1, & x \leq \zeta \\        
                \nu_2=\frac{\tan(1/3)\tan(20/3)}{10}, & x > \zeta
              \end{cases},\\
    a(x) &= \begin{cases}
                a_1=0.01, & x \leq \zeta \\
                a_2=\frac{1}{10\tan(1/3)\tan(20/3)}, & x > \zeta
              \end{cases},\\
    \kappa(x) &= \begin{cases}
                \kappa_1=1, & x \leq \zeta \\
                \kappa_2=\frac{1}{10\tan(1/3)\tan(20/3)}, & x > \zeta
              \end{cases}.
\end{aligned}
\end{equation}

\begin{equation}\label{prob2_analsol}
\begin{aligned}
    p(x,t) &= \begin{cases}
                \cos\left(\frac{20}{3}\right)\sin\left(\frac{x}{3}\right)e^{-100t/101}, & x\leq\zeta \\
                \sin\left(\frac{1}{3}\right)\cos(10(1-x))e^{-100t/101}, & x>\zeta
              \end{cases},\\
    u(x,t) &= \begin{cases}
                -3\cos\left(\frac{20}{3}\right)\cos(x)e^{-100t/101}, & x \leq \zeta \\
                -\frac{\sin(1/3)\sin(10(1-x))}{\tan(1/3)\tan(20/3)}e^{-100t/101}, & x > \zeta
              \end{cases},
\end{aligned}
\end{equation}

Once again, the performance of the two I-PINNs frameworks is compared with each other, as well as with PINNs using a relatively simple architecture with identical hyperparameters for the three, as listed in Table~\ref{prob2_table}. It is interesting to note that, in this case, both I-PINNs with CoNN and I-PINNs with SNN approximates the solution with an RMSE of $\mathcal{O}(10^{-3})$ for both displacements and pressures. Conversely, PINNs struggle to capture the heterogeneous response, resulting in substantially higher RMSEs of $\mathcal{O}(10^{0})$ and $\mathcal{O}(10^{-1})$ for the two fields, respectively. Once again, it is noted that despite having fewer parameters, SNN requires more computation time (higher by 5\%) than CoNN.

\begin{table}[hbt!]
\centering
\caption{Hyper-parameters used in training both PINNs \& I-PINNs (CoNN and SNN) models for approximating the solution to the compressible fluid problem. The activation functions are hyperbolic-tangent (tanh) and Gaussian error linear unit (GELU).}
\label{prob2_table}
{\footnotesize
\begin{tabular}{p{4cm}|c|c|c}
    \hline
   \textbf{Parameter} & \textbf{I-PINNs} & \textbf{I-PINNs} & \textbf{PINNs} \\
   & \textbf{(CoNN)} & \textbf{(SNN)} \\
   \hline
   Layer architecture & [2,40,40,40,1]$\times$2 & [2,40,40,40,2] & [2,40,40,40,1]$\times$2 \\
   \hline
   Activation functions & GELU-tanh & GELU-tanh & GELU \\
   \hline
   Number of training points [domain interior, external boundaries, interface] & [4900,140,70] & [4900,200,100] & [4900,200,100] \\
   \hline
   Penalty [interface conditions, boundary conditions, initial condition] & [200, 500, 300] & [200, 500, 300] & [200, 500, 300] \\
   \hline
   Maximum iterations & 70000 & 70000 & 70000 \\
   \hline
   RMSE (displacements) & 5.7$\times 10^{-3}$ & 7.6$\times 10^{-3}$ & 1.9$\times 10^{0}$ \\
   \hline
   RMSE (pressures) & 4.5$\times 10^{-3}$ & 3.3$\times 10^{-3}$ & 3.0$\times 10^{-1}$ \\
   \hline
   Cost & 1.0 & 1.2 & 0.7 \\
   \hline
\end{tabular}
}
\end{table}

\begin{figure}[!hbt]
    \centering
    \includegraphics[width=1.0\textwidth]{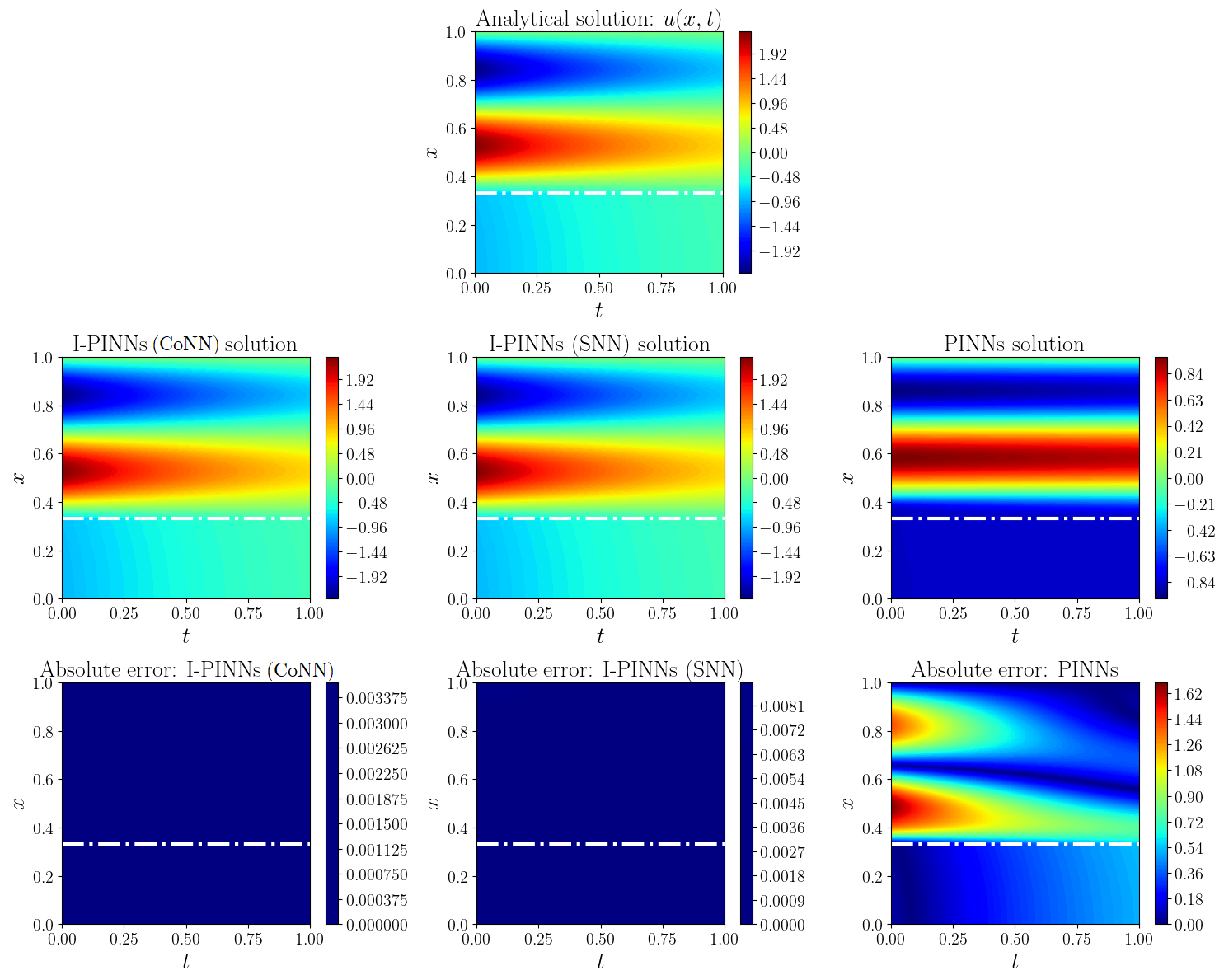}
    \caption{Displacement fields obtained by the two I-PINNs frameworks – SNN and CoNN, and as well as conventional PINNs (second row) for the compressible fluid problem ($a\neq0$). The third row displays the corresponding absolute errors. The interface (at $x=\zeta=1/3$) is demarcated with a white dash-dot line.}
    \label{prob2_u}
\end{figure}

\begin{figure}[!hbt]
    \centering
    \includegraphics[width=1.0\textwidth]{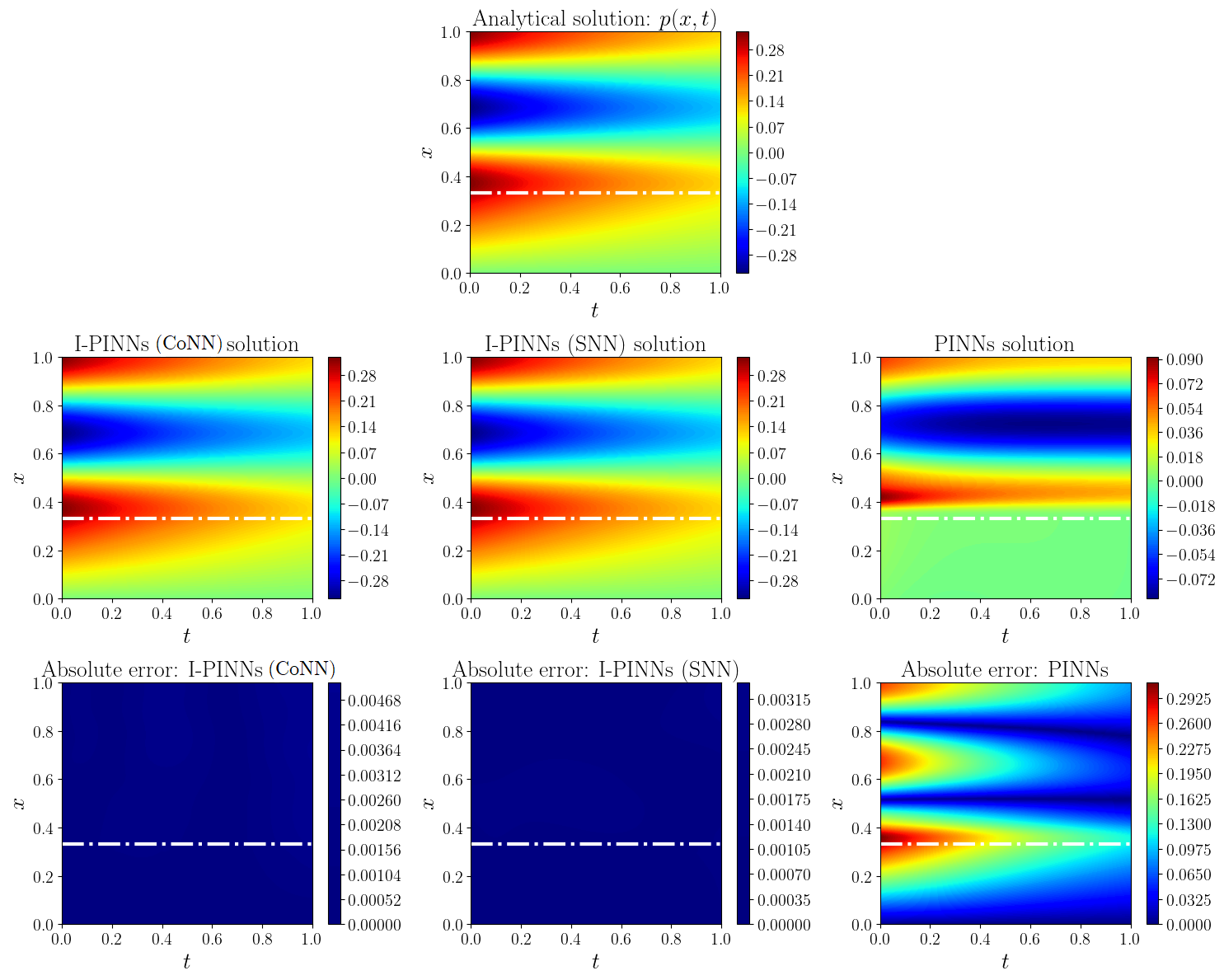}
    \caption{Pressure fields obtained by the two I-PINNs frameworks – SNN and CoNN, and as well as conventional PINNs (second row) for the compressible fluid problem ($a\neq0$). The third row displays the corresponding absolute errors. The interface (at $x=\zeta=1/3$) is demarcated with a white dash-dot line.}
    \label{prob2_p}
\end{figure}

Unlike the incompressible fluid case, here, the temporal derivative of pressure does not vanish ($\frac{\partial p}{\partial t}\neq 0$ as $a\neq0$). Therefore, displacements and pressures both exhibit strong transient characteristics due to their high coupling. The results for both displacements and pressures obtained from I-PINNs and PINNs are depicted in Figs.~\ref{prob2_u} and ~\ref{prob2_p}. Notably, in this case, rather than accumulating predominantly near the interface, the error is dispersed throughout the entire domain, particularly pronounced in $\Omega_2$. This error distribution signifies an inadequate approximation across the entire domain and demonstrates the limitations of conventional PINNs for transient poroelasticity in heterogeneous materials. Both the I-PINNs frameworks approximate the output fields well, with errors consistently falling within the same order of magnitude, i.e., $\mathcal{O}(10^{-3})$. Further insights from Figs.~\ref{prob2_u} and~\ref{prob2_p} reveal that, although the maximum error in displacements approximated by SNN is slightly higher than that of CoNN, the reverse is true for approximations in pressures, but this difference is very minor and negligible. This leads to the conclusion that, while for an incompressible fluid problem I-PINNs with CoNN outperforms I-PINNs with SNN, for a compressible fluid problem, they largely display the same level of accuracy in approximations. In the next section, the convergence profiles of the two frameworks are compared, and further insights into their performance are drawn.

\subsection{Convergence Behaviour}

\begin{figure}[!hbt]
    \centering
    \includegraphics[width=0.9\textwidth]{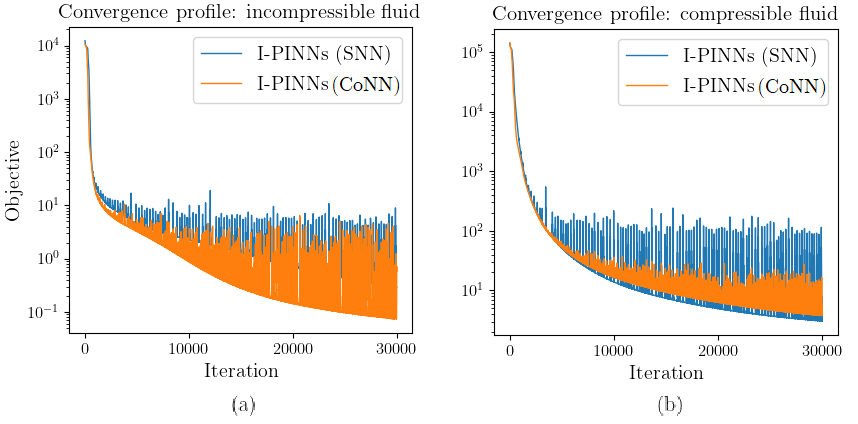}
    \caption{Convergence profiles depicting the optimization process for two experiments: (a) incompressible fluid and (b) compressible fluid. The plots display the objective function against iteration count for two models—I-PINNs with single neural network (SNN) and I-PINNs with composite neural network (CoNN).}
    \label{convergence_behaviour}
\end{figure}

Examining the convergence profiles in Figure~\ref{convergence_behaviour} for the two problems addressed by the I-PINNs frameworks, distinct observations emerge. In the context of the incompressible fluid problem, both frameworks—whether utilizing CoNN or an SNN—display a comparable level of oscillations during convergence. However, a notable difference is observed in the objective function values at each iteration, with CoNN consistently yielding significantly lower values than SNN. For the compressible fluid test case discussed in Section 4.2, the convergence profile for SNN exhibits noticeably higher oscillations compared to CoNN. This disparity highlights differing convergence dynamics between the two architectures, with CoNN demonstrating better convergence behavior for both test cases.

\subsection{Improvements on the Framework}
\label{sec:improvement}
Now, the I-PINNs (CoNN) framework is enhanced with the modifications outlined in section 3.3 (Glorot initialization, hard enforcement of boundary and initial conditions). The layer architecture is maintained consistently with the experiments in the previous sections (4 hidden layers with 40 neurons each), trained for 70,000 iterations. When hard enforcing the initial and boundary conditions, it is important to note that the number of points required for training decreases depending on which constraint is being hard enforced. Table~\ref{table_improvements} presents the RMSE when each of the modifications is applied to the vanilla version of our framework. 

Right from the outset, it is evident that the hard enforcement of boundary and initial conditions significantly improves the accuracy of the approximations. This improvement is particularly pronounced for the compressible fluid case, where enforcing these conditions effectively reduces the RMSE from $\mathcal{O}(10^{-3})$ to $\mathcal{O}(10^{-4})$ for both displacements and pressures. In the incompressible fluid scenario, a similar substantial improvement in pressures is observed, with both hard enforcement of initial conditions and boundary conditions resulting in a one-order-of-magnitude reduction in RMSE. However, for displacements, the significant decrease in RMSE from $\mathcal{O}(10^{-3})$ to $\mathcal{O}(10^{-4})$ is exclusively attributed to the hard enforcement of initial conditions while the impact of hard enforcement of boundary conditions remains negligible. The Glorot scheme for initializing network parameters (weights and biases) does contribute to improvements in the approximations, albeit marginally. While the order of magnitude remains consistent, there is a subtle reduction in magnitude.

\begin{table}[!hbt]
\caption{Performance comparison of the proposed I-PINNs with CoNN architectures with various modifications for the incompressible and compressible fluid flow problems. The accuracy is measured using the root mean square error (RMSE) metric.}
\centering
{\footnotesize
\begin{tabular}{c|c|c|c|c}
    \hline
     & \textbf{I-PINNs} & + \textbf{Glorot} & \textbf{+ Hard BC} & \textbf{+ Hard IC} \\
     & \textbf{(CoNN)} & \textbf{initialization} & & \\
     \hline
    Incompressible & 2.6$\times 10^{-4}$ & 1.5$\times 10^{-4}$ & 1.1$\times 10^{-4}$ & 5.6$\times 10^{-5}$\\
    (displacements) & & & & \\
    \hline
    Incompressible & 4.2$\times 10^{-4}$ & 3.5$\times 10^{-4}$ & 9.6$\times 10^{-5}$ & 3.1$\times 10^{-5}$\\
    (pressures) & & & & \\
    \hline
    Compressible & 5.7$\times 10^{-3}$ & 2.2$\times 10^{-3}$ & 7.6$\times 10^{-4}$ & 5.1$\times 10^{-4}$\\
    (displacements) & & & & \\
    \hline
    Compressible & 4.5$\times 10^{-3}$ & 3.7$\times 10^{-3}$ & 8.3$\times 10^{-4}$ & 6.5$\times 10^{-4}$\\
    (pressures) & & & & \\
    \hline
\end{tabular}
}
\label{table_improvements}
\end{table}

\section{Comparison with eXtended PINNs (XPINNs)}
\label{sec:comparison_with_xpinns}

In this section, a series of 1D experiments are conducted (using the same problems as in the previous section) to compare the performance of the proposed method with a widely used domain-decomposition-based PINNs framework known as eXtended PINNs (XPINNs)~\cite{jagtap2020extended} in terms of computational accuracy and cost. To ensure a fair comparison, the hyperparameters for both methods were selected to be as similar as possible. It is important to note that the key distinction between I-PINNs and XPINNs lies in their approach to subdomain decomposition along the interface: I-PINNs share the same set of parameters across subdomains, while XPINNs employ a different set of parameters for each neural network in the subdomains. Additionally, the loss terms related to the interface for the two methods are different. The proposed method enforces interface compatibility constraints as per Eqs. 13(h)-13(k), whereas the XPINNs implementation enforces flux continuity and the average value of the jump in the primary variable at the interface, as described in ‘Remark 5’ of Jagtap et al.~\cite{jagtap2020extended}. Furthermore, to maintain a consistent bias–variance trade-off between the models, the number of training points in XPINNs was proportionately increased compared to I-PINNs

\subsection{Comparison 1: Heterogeneous Poroelastic Media with an Incompressible Fluid}
Firstly, the problem described in Section~\ref{subsec:incompressible_fluid_numerical} is revisited, which involves incompressible fluid flow through a poroelastic medium with a material interface at $x=\zeta=1/5$, dividing the domain into two non-overlapping subdomains, $x_1\in(0,\zeta)$, and $x_2\in(\zeta,1)$. The two frameworks, I-PINNs–with–CoNN and XPINNs–with–CoNN, are applied using the hyperparameters listed in Table~\ref{prob1_table_ipinns_xpinns}. From the outset, it is evident that although XPINNs utilize approximately 25\% more parameters than I-PINNs, the root-mean-square error (RMSE) for both pressure and displacement approximations is of the order $\mathcal{O}(10^{-2})$, which is one order higher than that of I-PINNs ($\mathcal{O}(10^{-3})$ for both). Additionally, for this particular example, the computational cost of XPINNs is 40\% higher than that of I-PINNs, indicating a direct relationship between computational cost and the number of trainable parameters.

\begin{table}[hbt!]
\centering
\caption{Hyper-parameters used in training I-PINNs and XPINNs models for approximating the solution to the incompressible fluid problem.}
\label{prob1_table_ipinns_xpinns}
{\footnotesize
\begin{tabular}{p{4cm}|c|c}
    \hline
   \textbf{Parameter} & \textbf{I-PINNs (CoNN)} & \textbf{XPINNs (CoNN)} \\
   \hline
   Layer architecture & [2,20,20,20,1]$\times$2 & [2,15,15,15,2]$\times$2 \\
   \hline
   Activation functions & GELU-tanh & GELU-tanh \\
   \hline
   Number of training points [domain interior, external boundaries, interface] & [4900,140,70] & [5300,180,95] \\
   \hline
   Total \# of parameters & 1842 & 2164 \\
   \hline
   Maximum iterations & 20000 & 20000 \\
   \hline
   RMSE (displacements) & 3.8$\times 10^{-3}$ & 9.2$\times 10^{-1}$ \\
   \hline
   RMSE (pressures) & 3.6$\times 10^{-3}$ & 1.5$\times 10^{-2}$ \\
   \hline
   Cost & 1 & 1.5 \\
   \hline
\end{tabular}
}
\end{table}

\subsection{Comparison 2: Heterogeneous Poroelastic Media with a Compressible Fluid}
In the second experiment, the problem defined in Section~\ref{subsec:compressible_fluid_numerical} is considered, involving compressible fluid flow within a poroelastic medium with a material interface at $x=\zeta=1/3$. Both I-PINNs–with–CoNN and XPINNs–with–CoNN are utilized to approximate the solution, using the listed in Table~\ref{prob3_table_ipinns_xpinns}. Similar to the previous example, XPINNs produced the greater RMSE of the two, with pressure errors being one order of magnitude higher and displacement errors being two orders of magnitude higher than those of I-PINNs. While these differences might appear minor in simple academic problems, they could become more significant in complex practical applications, especially those involving multiple interfaces or inverse problems.

\begin{table}[hbt!]
\centering
\caption{Hyper-parameters used in training I-PINNs and XPINNs models for approximating the solution to the compressible fluid problem.}
\label{prob3_table_ipinns_xpinns}
{\footnotesize
\begin{tabular}{p{4cm}|c|c}
    \hline
   \textbf{Parameter} & \textbf{I-PINNs (CoNN)} & \textbf{XPINNs (CoNN)} \\
   \hline
   Layer architecture & [2,20,20,20,1]$\times$2 & [2,15,15,15,2]$\times$2 \\
   \hline
   Activation functions & GELU-tanh & GELU-tanh \\
   \hline
   Number of training points [domain interior, external boundaries, interface] & [4900,140,70] & [5300,180,95] \\
   \hline
   Total \# of parameters & 1842 & 2164 \\
   \hline
   Maximum iterations & 20000 & 20000 \\
   \hline
   RMSE (displacements) & 7.4$\times 10^{-3}$ & 1.4$\times 10^{-2}$ \\
   \hline
   RMSE (pressures) & 8.1$\times 10^{-3}$ & 1.1$\times 10^{-2}$ \\
   \hline
   Cost & 1 & 1.4 \\
   \hline
\end{tabular}
}
\end{table}

\section{Discussions}
\label{sec:discussions}

While the approach presented in this manuscript has demonstrated promise in addressing coupled-field problems such as poroelasticity in heterogeneous media, several improvements to the approach are necessary before it can be applied to model field-scale applications. For instance, the numerical examples and comparative studies presented were restricted to 1D cases with material coefficient mismatches of $\mathcal{O}(10^{1})$. Looking ahead, the extension of the proposed method to two-dimensional and three-dimensional poroelastic problems, with complex topologies and diverse interface shapes, is of interest and will be the subject of future research. Furthermore, applications involving moving interfaces and greater mismatches in material parameters are also of significant interest and will be explored in subsequent studies. Moreover, the hyperparameters were selected using a manual grid search method; however, it is recommended that practitioners employ automatic hyperparameter optimization techniques, such as Bayesian optimization, to identify the optimal parameters and further enhance the framework's performance. Although causality was enforced using a biased sampling approach, more advanced techniques like the causal training approach~\cite{wang2024respecting}, or a time-window approach with exact enforcement of temporal continuity should be used for long-term prediction of time-dependent problems~\cite{roy2024exact}.

\section{Conclusions}
\label{sec:conclusions}

In this study, a novel Physics-Informed Neural Networks (PINNs) framework is introduced for modelling poroelasticity in heterogeneous media with material interfaces. The approach incorporates two modifications to the conventional PINNs architecture. Firstly, a composite neural network (CoNN) is employed, where each sub-domain features a dedicated neural network for each output field variable (displacements and pressures). These neural networks utilize the same activation functions but are trained separately for their network parameters (weights and biases). Secondly, the neural network is integrated into the Interface-PINNs or I-PINNs framework (Sarma et al. 2024;
https://dx.doi.org/10.1016/j.cma.2024.117135), employing different activation functions across material interfaces to capture discontinuities in solution fields and gradients accurately. A single neural network (SNN) architecture integrated into the I-PINNs framework is also explored and compared against the proposed CoNN with respect to its accuracy and cost through two specific numerical examples involving an incompressible fluid and a compressible fluid. 

For both examples, the root-mean-square-error (RMSE) of conventional PINNs is found to be at least two orders of magnitude worse than that of CoNN and SNN. Although CoNN and SNN yield comparable accuracies, the latter is $20\%$ more expensive than the former in their run-time. Moreover, a comparison is performed between I-PINNs (with CoNN) with another popular domain-decomposition method called eXtended PINNs (XPINNs (with CoNN)), for the same set of problems discussed above. The results demonstrate that XPINNs (with CoNN), despite being at least 40 percent more expensive in their runtimes, yield RMSEs at least one order of magnitude worse than I-PINNs (with CoNN).

\section*{Acknowledgement}
\sloppy
Pratanu Roy’s work was performed under the auspices of the U.S. Department of Energy by Lawrence Livermore National Laboratory under Contract DE-AC52-07NA27344. Chandrasekhar Annavarapu gratefully acknowledges the support from ExxonMobil Corporation to the Subsurface Mechanics and Geo-Energy Laboratory under the grant SP22230020CEEXXU008957. The support from the Ministry of Education, Government of India and IIT Madras under the grant SB20210856CEMHRD008957 is also gratefully acknowledged.

\bibliographystyle{elsarticle-num} 
\bibliography{references}

\begin{thebibliography}{10}
\expandafter\ifx\csname url\endcsname\relax
  \def\url#1{\texttt{#1}}\fi
\expandafter\ifx\csname urlprefix\endcsname\relax\def\urlprefix{URL }\fi
\expandafter\ifx\csname href\endcsname\relax
  \def\href#1#2{#2} \def\path#1{#1}\fi

\bibitem{fielding1987application}
C.~R. Fielding, R.~C. Crane, An application of statistical modelling to the prediction of hydrocarbon recovery factors in fluvial reservoir sequences (1987).

\bibitem{morris2009injection}
J.~P. Morris, W.~W. McNab, S.~K. Carroll, Y.~Hao, W.~Foxall, J.~L. Wagoner, Injection and reservoir hazard management: the role of injection-induced mechanical deformation and geochemical alteration at in salah co2 storage project: status report quarter end, june 2009, in: LLNL Technical Report, 2009.

\bibitem{bloemendal2014achieve}
M.~Bloemendal, T.~Olsthoorn, F.~Boons, How to achieve optimal and sustainable use of the subsurface for aquifer thermal energy storage, Energy Policy 66 (2014) 104--114.

\bibitem{settari1998coupled}
A.~Settari, F.~Mounts, A coupled reservoir and geomechanical simulation system, Spe Journal 3~(03) (1998) 219--226.

\bibitem{menin2008mechanism}
A.~Menin, V.~A. Salomoni, R.~Santagiuliana, L.~Simoni, A.~Gens, B.~A. Schrefler, A mechanism contributing to subsidence above gas reservoirs and its application to a case study, International Journal for Computational Methods in Engineering Science and Mechanics 9~(5) (2008) 270--287.

\bibitem{lecampion2018numerical}
B.~Lecampion, A.~Bunger, X.~Zhang, Numerical methods for hydraulic fracture propagation: A review of recent trends, Journal of natural gas science and engineering 49 (2018) 66--83.

\bibitem{ammosov2022generalized}
D.~Ammosov, M.~Vasilyeva, E.~T. Chung, Generalized multiscale finite element method for thermoporoelasticity problems in heterogeneous and fractured media, Journal of Computational and Applied Mathematics 407 (2022) 113995.

\bibitem{adia2021combined}
J.-L. Adia, J.~Yvonnet, Q.-C. He, N.~Tran, J.~Sanahuja, A combined lattice-boltzmann-finite element approach to modeling unsaturated poroelastic behavior of heterogeneous media, Journal of Computational Physics 437 (2021) 110334.

\bibitem{sokolova2019multiscale}
I.~Sokolova, M.~G. Bastisya, H.~Hajibeygi, Multiscale finite volume method for finite-volume-based simulation of poroelasticity, Journal of Computational Physics 379 (2019) 309--324.

\bibitem{ahmad2022numerical}
Q.~A. Ahmad, M.~I. Ehsan, N.~Khan, A.~Majeed, A.~Zeeshan, R.~Ahmad, F.~M. Noori, Numerical simulation and modeling of a poroelastic media for detection and discrimination of geo-fluids using finite difference method, Alexandria Engineering Journal 61~(5) (2022) 3447--3462.

\bibitem{valiveti2023grid}
D.~M. Valiveti, C.~A. Srinivas, V.~Dyadechko, Grid modification during simulated fracture propagation, ~US Patent No. 11,608,730 (Mar.~21 2023).

\bibitem{raissi2019physics}
M.~Raissi, P.~Perdikaris, G.~E. Karniadakis, Physics-informed neural networks: A deep learning framework for solving forward and inverse problems involving nonlinear partial differential equations, Journal of Computational physics 378 (2019) 686--707.

\bibitem{haghighat2021physics}
E.~Haghighat, M.~Raissi, A.~Moure, H.~Gomez, R.~Juanes, A physics-informed deep learning framework for inversion and surrogate modeling in solid mechanics, Computer Methods in Applied Mechanics and Engineering 379 (2021) 113741.

\bibitem{zhang2022analyses}
E.~Zhang, M.~Dao, G.~E. Karniadakis, S.~Suresh, Analyses of internal structures and defects in materials using physics-informed neural networks, Science advances 8~(7) (2022) eabk0644.

\bibitem{cai2021physicsfluid}
S.~Cai, Z.~Mao, Z.~Wang, M.~Yin, G.~E. Karniadakis, Physics-informed neural networks (pinns) for fluid mechanics: A review, Acta Mechanica Sinica 37~(12) (2021) 1727--1738.

\bibitem{wessels2020neural}
H.~Wessels, C.~Wei{\ss}enfels, P.~Wriggers, The neural particle method--an updated lagrangian physics informed neural network for computational fluid dynamics, Computer Methods in Applied Mechanics and Engineering 368 (2020) 113127.

\bibitem{cai2021physics}
S.~Cai, Z.~Wang, S.~Wang, P.~Perdikaris, G.~E. Karniadakis, Physics-informed neural networks for heat transfer problems, Journal of Heat Transfer 143~(6) (2021) 060801.

\bibitem{jalili2024physics}
D.~Jalili, S.~Jang, M.~Jadidi, G.~Giustini, A.~Keshmiri, Y.~Mahmoudi, Physics-informed neural networks for heat transfer prediction in two-phase flows, International Journal of Heat and Mass Transfer 221 (2024) 125089.

\bibitem{sarkar1ause}
D.~R. Sarkar, C.~Annavarapu, P.~Roy, On the use of physics-informed neural networks to solve inverse problems in heterogeneous materials.

\bibitem{chen2020physics}
Y.~Chen, L.~Lu, G.~E. Karniadakis, L.~Dal~Negro, Physics-informed neural networks for inverse problems in nano-optics and metamaterials, Optics express 28~(8) (2020) 11618--11633.

\bibitem{haghighat2022physics}
E.~Haghighat, D.~Amini, R.~Juanes, Physics-informed neural network simulation of multiphase poroelasticity using stress-split sequential training, Computer Methods in Applied Mechanics and Engineering 397 (2022) 115141.

\bibitem{bekele2020physics}
Y.~W. Bekele, Physics-informed deep learning for flow and deformation in poroelastic media, arXiv preprint arXiv:2010.15426 (2020).

\bibitem{bekele2024physics}
Y.~W. Bekele, Physics-informed neural networks with curriculum training for poroelastic flow and deformation processes, arXiv preprint arXiv:2404.13909 (2024).

\bibitem{millevoi4074416physics}
C.~Millevoi, N.~Spiezia, M.~Ferronato, On physics-informed neural networks architecture for coupled hydro-poromechanical problems, Available at SSRN 4074416.

\bibitem{sarma2024interface}
A.~K. Sarma, S.~Roy, C.~Annavarapu, P.~Roy, S.~Jagannathan, Interface pinns (i-pinns): A physics-informed neural networks framework for interface problems, Computer Methods in Applied Mechanics and Engineering 429 (2024) 117135.

\bibitem{jagtap2020extended}
A.~D. Jagtap, G.~E. Karniadakis, Extended physics-informed neural networks (xpinns): A generalized space-time domain decomposition based deep learning framework for nonlinear partial differential equations, Communications in Computational Physics 28~(5) (2020).

\bibitem{coussy2004poromechanics}
O.~Coussy, Poromechanics, John Wiley \& Sons, 2004.

\bibitem{li2006support}
H.-s. Li, Z.-z. L{\"u}, Z.-f. Yue, Support vector machine for structural reliability analysis, Applied Mathematics and Mechanics 27~(10) (2006) 1295--1303.

\bibitem{ceryan2013modeling}
N.~Ceryan, U.~Okkan, P.~Samui, S.~Ceryan, Modeling of tensile strength of rocks materials based on support vector machines approaches, International Journal for Numerical and Analytical Methods in Geomechanics 37~(16) (2013) 2655--2670.

\bibitem{joshi2023machine}
S.~Joshi, S.~K. Singh, S.~Dubey, Machine learning and molecular dynamics based models to predict the temperature dependent elastic properties of silver nanowires, International Journal for Computational Methods in Engineering Science and Mechanics (2023) 1--9.

\bibitem{williams2006gaussian}
C.~K. Williams, C.~E. Rasmussen, Gaussian processes for machine learning, Vol.~2, MIT press Cambridge, MA, 2006.

\bibitem{mujumdar2007artificial}
A.~Mujumdar, P.~Robi, M.~Malik, M.~Horio, Artificial neural network (ann) model for prediction of mixing behavior of granular flows, International Journal for Computational Methods in Engineering Science and Mechanics 8~(3) (2007) 149--158.

\bibitem{hsiao2005fuzzy}
F.-H. Hsiao, W.-L. Chiang, C.-W. Chen, Fuzzy control for nonlinear systems via neural-network-based approach, International Journal for Computational Methods in Engineering Science and Mechanics 6~(3) (2005) 145--152.

\bibitem{al2012performance}
R.~Al-Ajmi, H.~Abou-Ziyan, M.~Mahmoud, Performance evaluation of 14 neural network architectures used for predicting heat transfer characteristics of engine oils, International Journal for Computational Methods in Engineering Science and Mechanics 13~(1) (2012) 60--75.

\bibitem{yagawa1996neural}
G.~Yagawa, H.~Okuda, Neural networks in computational mechanics, Archives of Computational Methods in Engineering 3 (1996) 435--512.

\bibitem{rall1981automatic}
L.~B. Rall, Automatic differentiation: Techniques and applications, Springer, 1981.

\bibitem{selmic2002neural}
R.~R. Selmic, F.~L. Lewis, Neural-network approximation of piecewise continuous functions: application to friction compensation, IEEE transactions on neural networks 13~(3) (2002) 745--751.

\bibitem{sarma2023variational}
A.~Sarma, C.~Annavarapu, P.~Roy, S.~Jagannathan, D.~Valiveti, Variational interface physics informed neural networks (vi-pinns) for heterogeneous subsurface systems, in: ARMA US Rock Mechanics/Geomechanics Symposium, ARMA, 2023, pp. ARMA--2023.

\bibitem{zhang2022multi}
B.~Zhang, G.~Wu, Y.~Gu, X.~Wang, F.~Wang, Multi-domain physics-informed neural network for solving forward and inverse problems of steady-state heat conduction in multilayer media, Physics of Fluids 34~(11) (2022).

\bibitem{roy2024adaptive}
S.~Roy, C.~Annavarapu, P.~Roy, A.~K. Sarma, Adaptive interface-pinns (adai-pinns): An efficient physics-informed neural networks framework for interface problems, arXiv preprint arXiv:2406.04626 (2024).

\bibitem{guo2022novel}
J.~Guo, H.~Wang, C.~Hou, A novel adaptive causal sampling method for physics-informed neural networks, arXiv preprint arXiv:2210.12914 (2022).

\bibitem{kingma2014adam}
D.~P. Kingma, J.~Ba, Adam: A method for stochastic optimization, arXiv preprint arXiv:1412.6980 (2014).

\bibitem{jax2018github}
J.~Bradbury, R.~Frostig, P.~Hawkins, M.~J. Johnson, C.~Leary, D.~Maclaurin, G.~Necula, A.~Paszke, J.~Vander{P}las, S.~Wanderman-{M}ilne, Q.~Zhang, \href{http://github.com/google/jax}{{JAX}: composable transformations of {P}ython+{N}um{P}y programs} (2018).
\newline\urlprefix\url{http://github.com/google/jax}

\bibitem{bisong2019google}
E.~Bisong, E.~Bisong, Google colaboratory, Building machine learning and deep learning models on google cloud platform: a comprehensive guide for beginners (2019) 59--64.

\bibitem{chiu2022can}
P.-H. Chiu, J.~C. Wong, C.~Ooi, M.~H. Dao, Y.-S. Ong, Can-pinn: A fast physics-informed neural network based on coupled-automatic--numerical differentiation method, Computer Methods in Applied Mechanics and Engineering 395 (2022) 114909.

\bibitem{peng2022rpinns}
P.~Peng, J.~Pan, H.~Xu, X.~Feng, Rpinns: Rectified-physics informed neural networks for solving stationary partial differential equations, Computers \& Fluids 245 (2022) 105583.

\bibitem{wang2023expert}
S.~Wang, S.~Sankaran, H.~Wang, P.~Perdikaris, An expert's guide to training physics-informed neural networks, arXiv preprint arXiv:2308.08468 (2023).

\bibitem{glorot2010understanding}
X.~Glorot, Y.~Bengio, Understanding the difficulty of training deep feedforward neural networks, in: Proceedings of the thirteenth international conference on artificial intelligence and statistics, JMLR Workshop and Conference Proceedings, 2010, pp. 249--256.

\bibitem{deng2023physical}
J.~Deng, X.~Li, J.~Wu, S.~Zhang, W.~Li, Y.-G. Wang, Physical informed neural networks with soft and hard boundary constraints for solving advection-diffusion equations using fourier expansions, arXiv preprint arXiv:2306.12749 (2023).

\bibitem{berrone2022enforcing}
S.~Berrone, C.~Canuto, M.~Pintore, N.~Sukumar, Enforcing dirichlet boundary conditions in physics-informed neural networks and variational physics-informed neural networks, arXiv preprint arXiv:2210.14795 (2022).

\bibitem{liu2022unified}
S.~Liu, H.~Zhongkai, C.~Ying, H.~Su, J.~Zhu, Z.~Cheng, A unified hard-constraint framework for solving geometrically complex pdes, Advances in Neural Information Processing Systems 35 (2022) 20287--20299.

\bibitem{bean2014immersed}
M.~Bean, S.-Y. Yi, An immersed interface method for a 1d poroelasticity problem with discontinuous coefficients, Journal of computational and applied mathematics 272 (2014) 81--96.

\bibitem{wang2024respecting}
S.~Wang, S.~Sankaran, P.~Perdikaris, Respecting causality for training physics-informed neural networks, Computer Methods in Applied Mechanics and Engineering 421 (2024) 116813.

\bibitem{roy2024exact}
P.~Roy, S.~T. Castonguay, Exact enforcement of temporal continuity in sequential physics-informed neural networks, Computer Methods in Applied Mechanics and Engineering 430 (2024) 117197.

\end{thebibliography}

\end{document}